\documentclass{article}

\usepackage{arxiv}

\usepackage[utf8]{inputenc} 
\usepackage[T1]{fontenc}    
\usepackage{hyperref}       
\usepackage{url}            
\usepackage{booktabs}       
\usepackage{amsfonts}       
\usepackage{nicefrac}       
\usepackage{microtype}      
\usepackage{lipsum}
\usepackage{verbatim}
\usepackage{graphicx}
\usepackage{amsmath}
\usepackage[english]{babel}
\usepackage[authoryear]{natbib}
\DeclareMathOperator*{\argmax}{argmax}

\title{Deep generative models for musical audio synthesis}

\author{
  Muhammad Huzaifah\\
  NUS Graduate School for Integrative\\ Sciences and Engineering\\
  National University of Singapore\\
  Singapore \\
  \texttt{E0029863@u.nus.edu} \\
   \And
 Lonce Wyse \\
  Department of Communications and New Media\\
  National University of Singapore\\
   Singapore \\
  \texttt{lonce.wyse@nus.edu.sg} \\
}

\begin{document}
\maketitle

\begin{abstract}
Sound modelling is the process of developing algorithms that generate sound under parametric control. There are a few distinct approaches that have been developed historically including modelling the physics of sound production and propagation, assembling signal generating and processing elements to capture acoustic features, and manipulating collections of recorded audio samples. While each of these approaches has been able to achieve high-quality synthesis and interaction for specific applications, they are all labour-intensive and each comes with its own challenges for designing arbitrary control strategies. Recent generative deep learning systems for audio synthesis are able to learn models that can traverse arbitrary spaces of sound defined by the data they train on. Furthermore, machine learning systems are providing new techniques for designing control and navigation strategies for these models. This paper is a review of developments in deep learning that are changing the practice of sound modelling.  
\end{abstract}

\vspace{5mm}
{\small This is the authors’ own pre-submission version of a chapter for Handbook of Artificial Intelligence for Music: Foundations, Advanced Approaches, and Developments for Creativity, edited by Eduardo R. Miranda, for Springer.}

\section{Introduction}
\label{sec:intro}
The development of capable ``guided'' parametric synthesis systems, that is the generation of sound given some user-determined parametric inputs, remains as one of the foremost challenges in digital audio production. A central challenge is that it is not enough for a model to generate a particular sound whether it be a drop of water or a whole symphony. A sound model includes an interface that can be used to traverse a space of sounds along different paths. A musical wind instrument, for example, affords access to a constrained class of sounds, and allows fingering patterns for the control of pitch, embouchure for timbre, or breath pressure for loudness. There is no \textit{a priori} limit on the range of sounds a designer might want from a model, nor the trajectories through the space they might desire to navigate via the interface. Deep learning-based generative models can be seen as spanning ground between two types of control principles -- one in which the user directly controls all aspects of the synthesis at each step akin to playing an instrument, and another whereby almost all decisions are left to the system, allowing sound to be generated unconditionally. There exists a broad spread in research focus along these lines, from finding novel control strategies for pre-existing models to designing algorithms comprising the sound synthesis.

Although each extreme has its merit and use cases, there is an increasing need in media production for more adaptable systems that marry extensive modelled knowledge to the flexibility of shaping the audio output in real-time. For instance, a live performer may wish to blend the timbres of several distinct acoustic instruments with natural sounds to create unique soundscapes on the fly. On the other hand, a developer may want to procedurally generate background music in a computer game based on certain in-game cues. A new approach to modelling audio utilising deep learning paradigms may offer an avenue to build such systems.

Deep learning has seen a surge of interest in the recent past, not least because of its huge potential in actualising many practical applications, including many areas of signal processing. It is already in widespread use in the music information retrieval community \citep{choi2017tutorial}, while many have declared automatic speech recognition as a largely ``solved'' problem with the advent of deep learning technology \citep{asr2019}. The remarkable ability of deep learning models to extract semantically useful features and utilise them for such downstream tasks have directed researchers to not only purpose these models to analyse and process existing data, but to actually \textit{generate} new data. 

Deep generative models are a powerful subset of deep learning networks that discover latent structure in data in order to generate new samples with the same distribution as the training data. Unlike more common learning objectives that try to discriminate labelled inputs (i.e. classification) or estimate a mapping (i.e. regression), generative models instead learn to replicate the hidden statistics behind observed data. This ``understanding'' of the structure of data space allows them to display impressive expression capabilities on a variety of audio and non-audio-related tasks. For image media, state-of-the-art generative adversarial networks are now able to synthesize extremely life-like human faces, even retaining some control over both broad and fine-grained stylistic features in the process \citep{karras2019style}. Audio data has unique characteristics that make it a challenge to synthesize with reasonable fidelity using existing techniques derived primarily for visual applications. Despite this, current deep learning models have often shown to outperform previous widely used parametric models such as the hidden Markov model, especially in applications where adequate data is available. 

In the following sections we will analyse several key deep generative models developed for musical audio synthesis. This is prefaced by a discussion on the audio synthesis task and as well as a broader introduction to more generic generative neural networks that form the basis of the systems used for audio. Although the focus of this chapter will be on the musical domain, some discussion on speech models will also be included to provide a better general picture of the capabilities of deep generative audio models, and because there are many overlapping issues concerning audio synthesis and issues of designing real-time control. We will also give little attention to musical modelling in the symbolic domain of notes except where it is directly relevant to audio modelling such as when notes are used as parameters to conditionally synthesise audio.

\section{Overview}
\label{sec:overview}

\subsection{Problem background}
\label{sec:background}
Synthetically-generated media is ubiquitous today. This holds true in the musical domain, where digital synthesizers have been widely and increasingly used in both production and performance. Artists and engineers work with synthesizers in a variety of roles that typically fall within one of the following scenarios:
\begin{itemize}
\item generate interesting or novel sounds or timbres that are infeasible/impossible to be produced acoustically
\item simulate the sounds of real-world acoustic instruments or of other sources such as environmental sounds
\item facilitate the automation of systems and processes (e.g. text-to-speech, virtual assistants etc.)
\end{itemize}

\begin{figure}[h]
\begin{center}
\includegraphics[width=0.61\textwidth]{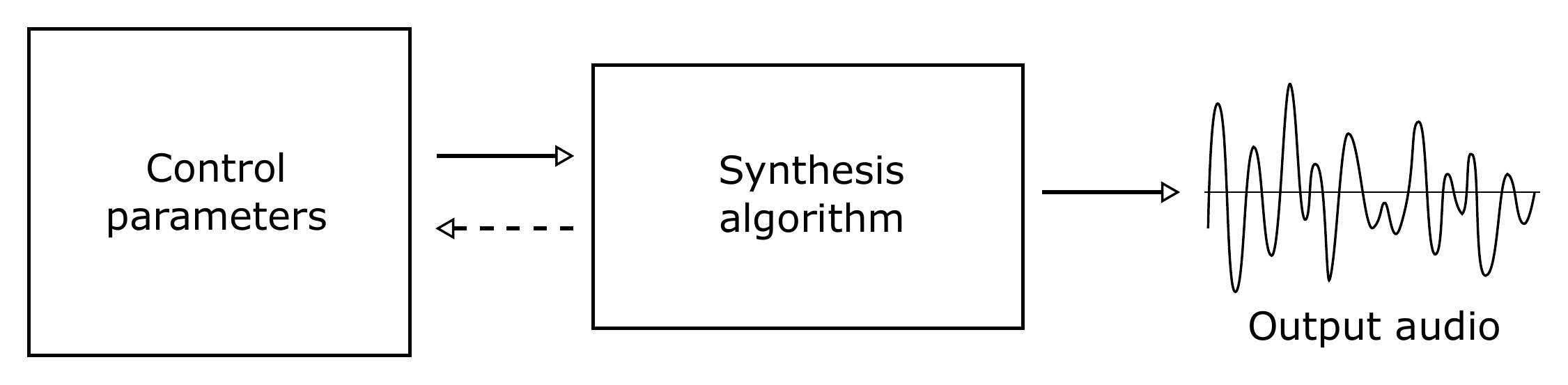}
\end{center}
\caption{A high-level system view of the audio generative task. A user interacts with the system via control parameters to change various qualities of the waveform. A synthesis algorithm determines what kind of parameters are available and their relation to the overall waveform, which would indirectly instruct how the user goes about conceiving a particular sound.}
\label{fig:synthchain}
\end{figure}

In terms of a computational task, the process driven by a digital synthesizer is that of guided audio synthesis. Succinctly, the aim is to produce a sound with particular characteristics defined by the user. As illustrated in Fig. \ref{fig:synthchain}, we can further distil this overall objective into two associated aspects: that of \textit{generation} and that of \textit{control}. The former concerns the algorithm behind the actual sound production, whilst the latter relates to how this sound can be shaped (hence ``guided''). A considerable body of literature on sound modelling has been devoted to mapping musical gestures and sensors through an interface to the parameters a signal-processing algorithm makes available \citep{Wanderly2000,Hunt2003}. 

In the natural physical world, a sound is generated as a result of a physical interaction between two or more objects called ``sources'' or ``excitors'', that is transmitted in the form of pressure waves through a medium. In the digital realm, a synthesis algorithm tries to replicate the target sounds without having or requiring physical sources. There are many approaches to sound modelling, and each has advantages and disadvantages for different classes of sounds, as well as in terms of expressiveness, control, and computational efficiency. A ``physical modeling'' approach mathematically models the physics that generate a target waveform (see for example \citep{Smith92}). An example is using equations that model travelling waves along strings attached to resonating bodies and excited by a bow being drawn across the string. One advantage of physical modelling is that the equations expose parameters that are intuitive for control such as bow pressure, string length and stiffness, and resonating body size. An ``acoustic modeling'' approach uses signal generators and processors for manipulating waveforms such as oscillators, modulators, filters, and waveshapers designed to generate acoustic features regardless of physical processes. Commercial synthesizers are comprised of such units that can be configured to generate a vast variety of sounds, but expose algorithm-specific parameters that may not be as manageable or intuitive as physical model parameters for control. Prerecorded sound samples are also used in several techniques and can obtain a natural quality difficult for purely synthetic approaches. Techniques in this category include wavetable synthesis for musical instruments which uses looping to extend note durations and layers for timbral variation. Concatenative synthesis, more common for speech, draws on a large database of very short snippets of sound to assemble the target audio. Prerecorded samples have the naturalness of the recorded audio, but present their own control challenges for achieving arbitrary sequences of sound samples. Manually constructing synthesis algorithms that cover a specific range of sounds under desired parametric trajectories and that capture the complexity of natural audio is a difficult and labour-intensive task.

\subsection{Data-driven parametric synthesis}

Statistical models first came to prominence in the 1980s with the hidden Markov model (HMM) that eventually dominated the fields of automatic speech recognition and generation. HMMs use a series of hidden states to represent non-stationary acoustic features and are often combined with a Gaussian mixture model (GMM) that admits frame-wise mappings beneficial for several speech generation tasks such as voice conversion and speech enhancement \citep{ling2015deep}. Statistical models learn their model parameters from data as opposed to expert design in the case of physical and acoustic models. The data-driven nature of statistical parametric models would make them similar in that aspect to sample-based techniques. Indeed, data-driven parametric synthesis systems posses many of the positive qualities of the prior approaches. The more modern successor to HMMs are deep generative models, also known as generative neural networks, that are more powerful and scale much better with data. They are also capable of ``end-to-end'' learning (without separate components for feature extraction, processing, and feature-to-audio synthesis) whereas HMM-based models operate on and produce engineered features not learned by the model.

The idea behind generative neural network models is to synthesise sounds according to the learned characteristics of an audio database. The generated sound would therefore be perceptually similar to, but not merely be reproductions of, the data the model was trained on. The underlying assumption here is that the training data (audio plus metadata) contains all the necessary information required to recreate the different dimensions of the desired sound(s) without the model requiring \textit{a priori} knowledge of those properties. During training, the network is tasked to find patterns, dependencies, and variations in the data. This process can be framed as an instance of the inverse problem, where we try to determine the causal factors to a set of observations. Whereas physical modelling synthesis finds an analytical solution that corresponds to a set of equations governing the sound production and propagation, generative neural networks provide a numerical solution with model parameters (i.e. the network weights) that may or may not correspond to real-world physical or perceptual parameters.   



\subsection{Control affordances}
\label{sec:affordance}

One drawback of using the physical, acoustic, and sample-based approaches to interactive sound model design is that the interface to the model is either the set of parameters inherent in the equations or algorithms used for synthesis, or else it is designed separately by mapping desired control gestures to the parameters exposed by a synthesis algorithm. However, the  affordances \citep{soegaard2012encyclopedia} for interaction  are a critical part of the expressive potential of a sound model designed for music. Machine learning approaches have been used for mapping gestures to the space of algorithm parameters.  \citet*{FelsHinton1993} described a neural network for mapping hand gestures to parameters of a speech synthesizer. \citet{FiebrinkThesis2011} developed the Wekinator for mapping arbitrary gestures to parameters of sound synthesis algorithms.  \citet{FriedFiebrink2013} use stacked autoencoders for reducing the dimensionality of physical gestures, images, and audio clips. \citet{Francoise2014} developed various strategies for mapping complete gestures to parameters of synthesizers. \citet{FascianiWyse2012} used machine learning to map vocal gestures to sound, and separately to map from sound to synthesizer parameters for generating sound.  \citet{gabrielli2017introducing} used a convolutional neural network to learn upwards of 50 ``micro-parameters'' of a physical model of a pipe-organ. These techniques show the value of various machine learning techniques for gesture mapping, but they all use predefined synthesis systems for the generation of sound or images, and are thus limited by the capabilities of the synthesis algorithms they learn to control. They do not support learning mappings “end to end” between gestures and arbitrary sound sequences.   

How can we design networks that not only generate data from distributions on which they have been trained, but do so under the type of intimate control that we require of musical instruments? Deep learning networks do not facilitate reworking their internal structure during the generative phase as a viable means of controlling the output. Several studies from both audio and vision fields therefore focus on the training phase instead, to determine the characteristics of the post-training generative phase. This has seen the use of specialised architectures (e.g. Siamese networks that learn two different genres of music from a common input of sheet music \citep{malik2017neural}), specific loss functions (e.g. first versus second order activation losses to abstract ``style'' and ``content'' from images \citep{gatys2015neural} or spectrograms \citep{audio2016style}) or curated training sets (e.g. adding or removing representative data to bias the model output a certain way  \citep{fiebrink2016machine}). However, there are several limitations with these approaches, especially in the context of an audio synthesizer in a production setting. Firstly, they are fundamentally not dynamic in real-time, requiring time-consuming re-training to alter the synthesis. They furthermore conflate many perceptual dimensions without allowing more fine-tuned control over each independently. Also they do not offer easy ways to add new control dimensions to existing ones.

An alternative strategy available for controllable generative models is \textit{conditioning}, whereby auxiliary information is given to the model in addition to the audio  samples themselves during training. If the generative phase is run without such information, samples are drawn from the model distribution based on the whole training set. For instance, for a network trained on sample vectors of speech without being augmented with phonemic information, the generation produces ``babble''. However, when conditioned with the augmented input information during training, the phonemic part of the input can be externally specified so that the audio samples are generated conditionally based on the phonemes presented in the input sequence. This technique has been used successfully in networks used for sequence generation such as WaveNet \citep{van2016wavenet} and Char2Wav \citep{sotelo2017char2wav}, a conditional extension of SampleRNN \citep{mehri2016samplernn}. Conditional models provide a means to directly influence the synthesis with more intuitive labels since the dimensions are chosen by the model developer. Choosing a set of conditional parameter values is comparable to restricting the output generation to a restricted region of the complete data space. It is also possible to use learning mechanisms to discover low-dimensional representations that ``code'' for different regions of the whole data space and that can then be used during generation. Models that incorporate conditioning now form a huge part of generative deep learning literature and the concept has played an important role in their success. Conditioning strategies are addressed in more detail below. 




\subsection{Data representation and hierarchy}
\label{sec:rep}

\begin{figure}[h]
\begin{center}
\includegraphics[width=0.67\textwidth]{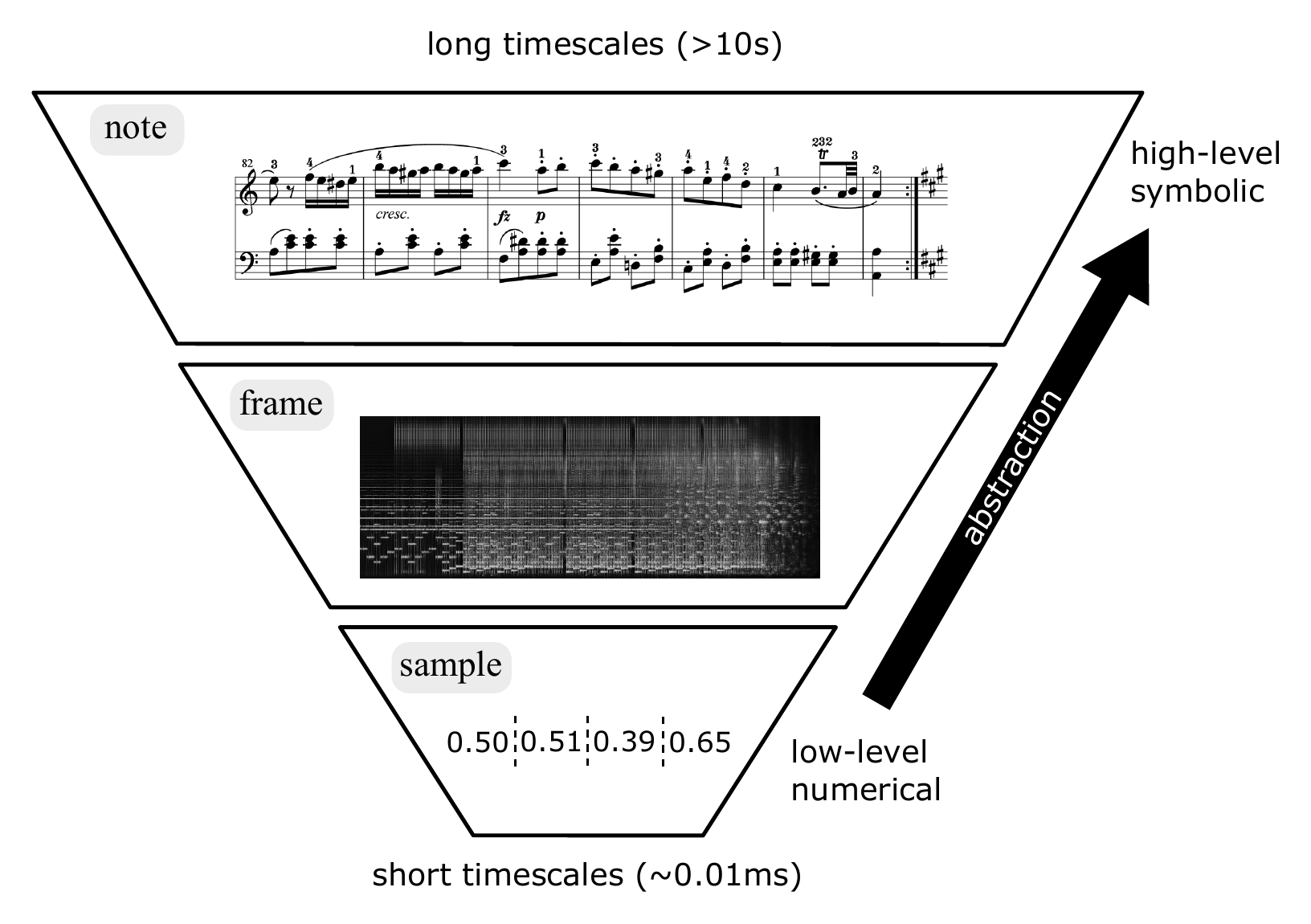}
\end{center}
\caption{Audio representation in the musical domain exists as a hierarchy of timescales and levels of abstraction, starting from smallest digital audio quantum of a sample, up to the entire musical passage, often represented symbolically in the form of a score. A generative model of audio that can capture structure all the different timescales may not be entirely feasible without compromising quality at some scale, usually manifesting as a loss in fidelity or a lack of long-term structure. The task can be made easier by focusing on a specific timescale and using the appropriate representation.}
\label{fig:hierarchy}
\end{figure}

At the heart of the deep generative model is the data used to train them. Audio as data is representationally challenging since it is a hierarchically structured medium that can be encoded in different forms. Moreover, it is multi-level, and can co-exist and experienced at many levels of abstraction (see Fig. \ref{fig:hierarchy}). What is representationally appropriate then is dependent upon the task objective, which would in turn greatly influence the model design. 

Many generative models of music are built around the use of high-level symbolic (e.g. note) representations such as MIDI, sheet music or piano rolls that abstract away the peculiarities of a particular musical performance. Although relatively simpler to model compared to other forms of audio, they lose the more fine-grained nuances that are present in performed music which may negatively impact the overall enjoyment of the listener. 

Another class of audio representations are transformed signals that represent ``frames'' or ``windows'' of audio samples in vectors that are sequenced in time at a much lower rate than sampled audio. Spectrograms are an example of a time versus frequency representation. A two-dimensional representation suggests the use of techniques developed for images for discriminative and generative tasks. However, audio data in the time and frequency dimensions behave very differently than typical image data, and don't necessarily suit the assumptions built in to image processing networks \citep{wyse2017audio,bin2019applying}. Furthermore, while a complex spectrogram can be lossless, a typical representation in practice is the magnitude-based spectrogram  which requires an audio signal to be reconstructed. The most common technique for ``inverting'' a spectrogram comes from \citet{GriffinLim84} but it requires information from across the duration of the entire signal, so is not directly applicable to real-time interactive generative systems, though more recent techniques have addressed this issue \citep{ZhuBeauWyse2007} and dramatically increased the quality \citep{Prusa2017} achievable. Other spectra-based representations such as mel-scaled spectrograms or mel-frequency cepstral coefficients (MFCCs) are common but even more lossy. Neural network ``vocoders'' have been developed that learn to generate plausible audio signals from such representations \citep{Tamamori2017}.     

The most straight-forward audio representation is the raw sampled audio waveform which is lossless and trivially convertible to actual sound. Unlike symbolic representations that may be instrument-specific, models that operate on raw audio samples can be applied to any set of instruments and even non-musical audio like speech and environmental sounds. All the musically relevant nuances and idiosyncrasies are also embedded in an audio waveform. On the other hand, modelling such waveforms is extremely challenging, particularly in the handling of the inherent long-term dependencies in sound over time. This is exacerbated by the fact that audio signals are sampled at high temporal resolutions, from between 16000 to 48000 samples per second. This makes capturing typical musical time dependencies that span many seconds or minutes difficult for some network architectures.  

As shown in Fig. \ref{fig:hierarchy}, the shift from shorter to longer timescales broadly coincides with the use of more symbolic representations to capture and transmit information at increasingly higher levels. The guided synthesis task of interest is, on the whole, more focused on local musical structures such as timbre and pitch. Accordingly, the systems that will be discussed here tend to directly model raw audio signals in the form of digital samples in a PCM stream or processed with other standard quantization encodings such as $\mu$-law. We do not address ``automatic music composition'', which concerns much longer timescales and attempts to model concepts that are relevant at the level of a song or musical piece like continuity, repetition, contrast, and harmony. The majority of automatic composition models also use symbolic representations as input data to abstract away low-level information, making the capture of longer time dependencies more feasible. PerformanceRNN \citep{oore2018time}, is an example of an automatic music composition network that jointly predicts MIDI notes and also their expressive timing and dynamics. 

In summary, the paradigm for guided audio synthesis put forth in this chapter focuses on modelling local structure and leave the longer-term evolution of the sound to user control. This entails working mainly in the raw audio domain which contains the greatest detail at short timescales.

\subsection{Audio domains}
\label{sec:domain}
The types of audio from the literature can essentially be divided into three spheres of speech, music, and environmental sounds. The primary concern here is with models for musical sound with a focus on pitch, timbre, and articulation characteristics. Nevertheless, a portion of the synthesis systems analysed were not developed for any particular domain, and like many deep learning models, are agnostic to the dataset used to train them. Since they are trained on raw audio, many of these generative models can be extended to music even though they were not specifically built for that purpose. A large number of generative models for audio were developed first and foremost for speech applications. Some speech models will also be included for a more comprehensive discussion on audio-based deep generative models as the development of the entire field runs parallel to the development of synthesis models for music.

\subsection{Related works}
\label{sec:related}
We end this section by providing a brief outline of other reviews of deep generative models for audio that may be of relevance. They mostly focus on specific audio domains. 

\citet{briot2017deep} provides a comprehensive survey of systems that generate musical content through a multi-pronged analysis in terms of objective, representation, architecture and strategy. Another work on music generation by deep learning \citep{briot2017music} instead orient their analysis in terms of domain challenges. They look at concepts such as originality, control, and interactivity, among others, and determine how current models can fill these gaps (or fall short of fully addressing the issues). For speech, \citet{ling2015deep} evaluates various statistical parametric models, noting the shift from the well-established hidden Markov model (HMM) and Gaussian mixture model (GMM) to deep learning architectures, though the systems highlighted mostly belong to older classes of deep learning models such as deep believe networks (DBNs) and restricted Boltzmann machines (RBMs). These nonetheless share similar theoretical underpinnings with the more modern autoencoders that are widely used for current generative tasks. \citet{henter2018deep} discusses unsupervised learning methods for expressive text-to-speech, especially in the context of variational autoencoders. \citet{DBLP:journals/corr/abs-1905-00078} goes against the trend of domain-specific surveys. They instead give a broad overview of the advances in deep learning as applied to audio processing. A wide range of methods, models, and applications including both analysis and synthesis aspects are covered albeit without going into detail. Perhaps the closest in terms of content to the review here is an overview of generative music models dealing with raw waveforms by \citet{waveformgenonline}, which, for the reader wanting to gain a more complete picture of the field, would serve as a good follow-up text to this work.







The next section will introduce some of the key deep learning architectures that form the basis for generative musical audio models, important for the understanding of how these models work and the motivation behind them. Specific generative systems from the literature will be discussed afterwards as part of a narrative on the conceptual development of a functional synthesizer.

\section{Generative neural networks}
\label{sec:gnn}
Statistical generative or predictive models based upon artificial neural networks have gained prominence since the development of the Generative Adversarial Network (GAN) in 2014. To better understand how they work, it is best to study generative models through the lens of probability theory. The core goal of a generative model is to approximate an underlying probability distribution $p_{data}$, given access to a finite set of samples drawn from this distribution. Training a generative model entails choosing the most plausible model parameters that minimise some notion of distance between the model distribution and the true data distribution. More formally, given training data points $X$ as samples from an empirical distribution $p_{data}(X)$, we want to learn a model $p_{\theta}(X)$, belonging to a model family $M$ that closely matches $p_{data}(X)$, by iteratively changing model parameters $\theta$. In short, we specify the following optimisation problem.

$$\min_{\theta \in M} d(p_{data},p_{\theta})$$

A commonly used objective function $d$ to evaluate the quality of the model and drive learning is the Kullback-Leibler (KL) divergence, a measurement of how different two distributions are, which satisfies the required notion of a quantifiable ``distance'' between the data distribution and model distribution. Equivalently, minimising the KL divergence can be thought of as maximising the log-likelihood of data points $X$ with respect to the model distribution $p_{\theta}$.

To understand just how difficult the problem of generating audio data is, we can have a look at the size of its state space, or the number of possible configurations a segment of audio can take. With a conservative assumption of 1 second of audio, sampled at 16kHz with 8 bits per sample (i.e. 256 possible values), there exists $256^{16000}\approx10^{38500}$ possible sequences, a number that will only get larger for longer sequences or with higher quality and therefore impossible to capture by just memorising all possible configurations. Fortunately, real world audio is highly structured, and it is this underlying structure that the generative neural network aims to learn.    

When trained in a supervised setting with inputs $X$ and targets $Y$, a generative model estimates the full joint probability $p(Y,X)$, which can not only predict $Y$ but also says something about the generative process of $X$. In contrast, a discriminative model would learn to output the most likely $Y$ for a given $X$ more directly with a conditional distribution\footnote{The relationship between conditional and joint probabilities can be seen by inspecting the chain rule: $p(Y,X) = p(Y|X)p(X)$. For classification, taking $\argmax p(Y|X)$ is equivalent to taking $\argmax p(Y,X)$ since $p(X)$ is constant. However the presence of the $p(X)$ term means that a generative model accommodates a more complete probabilistic description of the data compared to a discriminative model and hence capable of more than just prediction.} $p(Y|X)$. Although learning a conditional probability is oftentimes much easier, with the joint probability, a generative model can be used in a number of interesting ways. For our main objective of synthesizing audio, we can \textit{sample} from the model distribution to generate novel data points (in which case $Y$ is also a complex high-dimensional object like an audio waveform as opposed to a label). Since the model distribution is close to the dataset distribution, we get something perceptually similar to the training data. Generative models can also be used for \textit{density estimation}, that is assigning a likelihood value to a given $X$. This can provide an indication of how good the model actually is in approximating $p_{data}$. Finally, generative models are often used for \textit{unsupervised representation learning} as an intermediate objective. Internally, the deeper neural network layers making up the model have a significantly smaller number of parameters compared to the input data, thereby forcing them to discover and capture the essential dependencies mapping the low-level input to a higher-level abstracted representation known as the \textit{feature space/latent space}. The latent variables extracted by the model are often essential for further downstream inference tasks such as classification and conditional generation. With such flexible and powerful properties, researchers have used generative models for many objectives aside from synthesis, including data compression and data completion.         

\begin{figure}[t]
\begin{center}
\includegraphics[width=0.86\textwidth]{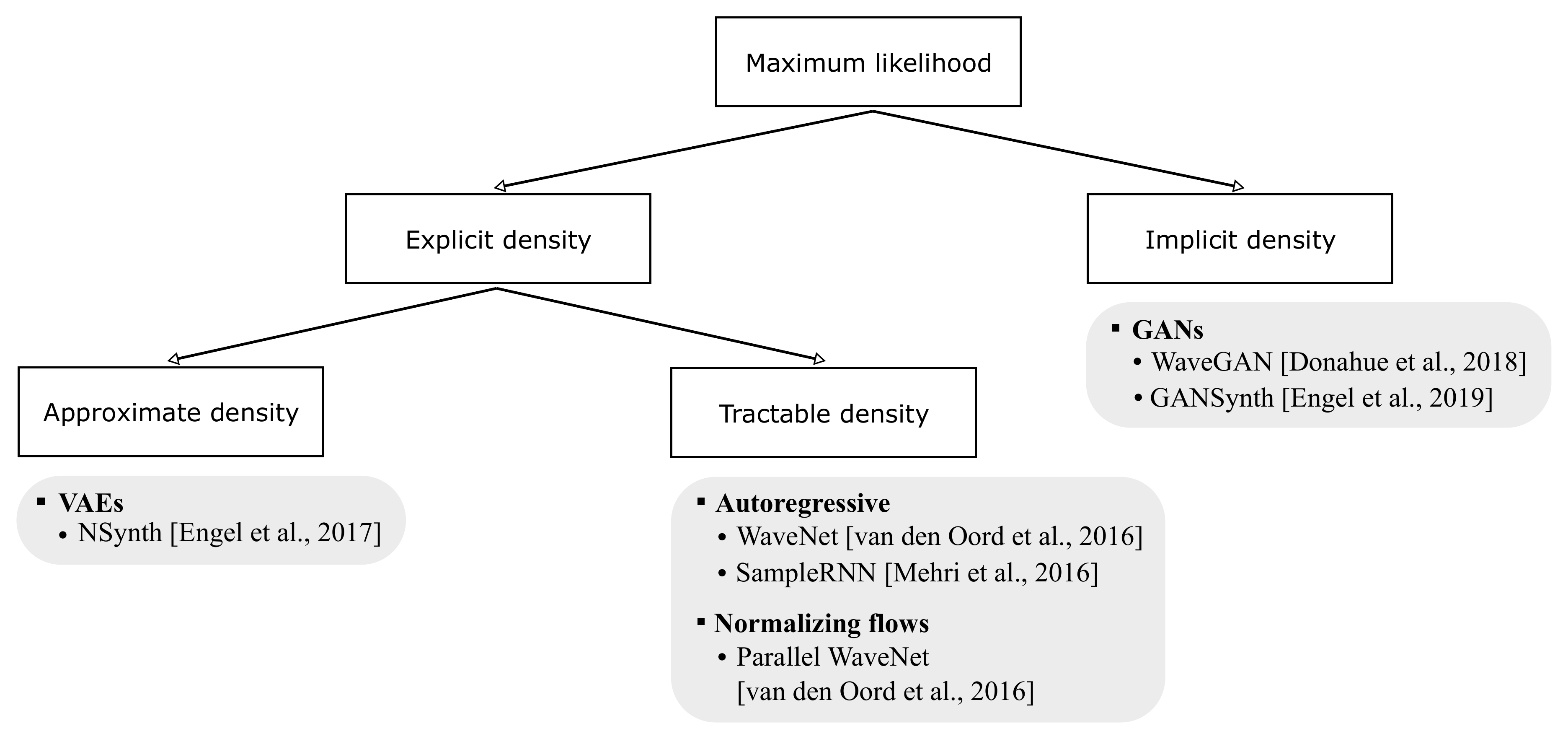}
\end{center}
\caption{One way of organising different generative models is by looking at how the model treats maximum likelihood and whether the density distribution is explicitly or implicitly expressed. This chapter covers some of the important generative model families in current use for audio synthesis. Autoregressive models (Sec.\ref{sec:autoregressive}), variational autoencoders (VAEs)(Sec.\ref{sec:VAE}) and normalizing flow models (Sec.\ref{sec:flow}) constitute explicit density models, while the generative adversarial network (GAN)(Sec.\ref{sec:GAN}) is an implicit density model. Included under each heading is a non-exhaustive list of essential audio-related works that utilise that model type.}
\label{fig:taxonomy}
\end{figure}

One caveat to take note of when using generative networks is that not all model families can perform equally well on every inference task. Indeed, the various trade-offs in their inference capabilities plus the particular assumptions made on a given dataset have lead to the development of a wide array of generative networks which we break down in Fig. \ref{fig:taxonomy}. Since many generative models are trained using maximum likelihood estimation as mentioned, a natural way to taxonomize generative models is in how they represent the likelihood. Some types of networks allow for the explicit formulation of the marginal likelihood $p(X)$ of the data. These include autoregressive networks and variational autoencoders. In other cases, a training regime specifies a stochastic process that instead requires the network to have implicit knowledge of the underlying probability distribution to sample from, an example of which is the GAN. In practice, being able to calculate the likelihood provides an unambiguous way of measuring how close the model is in approximating the real data distribution; with GANs, we can only evaluate the model by looking at the generated examples and comparing them to the real data.

The rest of this section will elaborate several of these model families that have been used for audio synthesis with varying success. Researchers have taken advantage of their different properties to fulfil a range of audio generative tasks and challenges. Generally from the most to least widespread in current use for audio they are:
\begin{itemize}
\item Autoregressive Models -- make sample predictions sequentially in time
\item Variational Autoencoders (VAEs) -- used to discover low-dimensional parametric representations of the data
\item Normalizing Flow Models -- for speeding up generation and modelling complicated distributions without ``bounding'' approximations
\item Generative Adversarial Networks (GANs) -- for speed and parallelization, developing representations that apply globally across the temporal extent of a sound, and producing large chunks of audio at a time
\end{itemize}


\subsection{Autoregressive Models}
\label{sec:autoregressive}
The first type of generative network in focus is designed specifically with sequential data in mind, and hence lends itself naturally to audio data viewed as a temporal sequence. Autoregressive networks define an explicit and computationally tractable density model by decomposing the probability distribution over an $n$-dimensional data space $X$ into a product of one-dimensional probability distributions via the chain rule of probability. 

$$p(X) = \prod_{i=1}^{n} p(x_i|x_1,...,x_{i-1})$$

We see that data is assumed to have a canonical sequential direction -- the current term in the sequence ($x_i$) is only conditioned on a recency window of previous terms and not on ``future'' terms relative to the current. These models learn to predict the next sample in time given what has come just prior.  It is in part this ``causality'' that permits real-time manipulation of the generated audio using techniques described below. Autoregressive models are also known to be easier to train in comparison to the other model families discussed after.   

The process of basing the subsequent prediction on previous terms may seem similar to the perhaps more familiar recurrent neural network (RNN). Indeed an RNN can be cast as a type of autoregressive model that compresses the prior terms into a hidden state instead providing them explicitly as input to the model. RNNs have been used as building blocks for more complex autoregressive models of audio such as SampleRNN \citep{mehri2016samplernn}. Each layer in SampleRNN is comprised of RNNs that work to capture dependencies at a particular timescale.

Autoregressive models however are not exclusive to the RNN. At the time of writing, the most prevalent generative model for audio was proposed by van den Oord et al. at DeepMind, known as WaveNet \citep{van2016wavenet}. Where an RNN sees only one input sample at each time step and retains the influence of past samples in its state, WaveNet has explicit access to a past window of input samples. The constituents of WaveNet are convolutional neural networks (CNNs) where each neural network layer learns multiple filters for which to process the input to that layer. To make CNNs abide by the autoregressive principle, the filters are partly masked to avoid a computation with non-causal inputs of the data. A big advantage of CNNs over RNNs is their capacity for parallelism, where a long input sequence can be processed quickly as a whole. This can greatly speed up training since the entire output can be processed in one forward pass. At generation time however, there is no training data to base previous time steps, so the model has to wait for each sample to be generated in turn in order to use it to predict the next sample. Slow generation is one inherent weakness of autoregressive networks that has been addressed by other model families, namely normalizing flows and GANs.   

Another drawback of autoregressive models is that they do not learn unsupervised representations of data directly by themselves, and so we do not have access to a feature space intrinsically. A workaround is combining an autoregressive model with a separate encoder to allow the autoregressive network to condition itself on latent variables provided by the encoder to augment the input data. \citet{van2016conditional} demonstrates other types of conditioning as well, including one-hot encodings of class labels and features taken from the top layers of a pre-trained CNN. The combination of an autoregressive model with an additional encoder network has been used extensively in the literature to impose some conditioning influence over the synthesis procedure, including for extensions to WaveNet as in \citet{engel2017neural}, discussed in more detail in Sec. \ref{sec:latent}.

\subsection{Variational Autoencoders}
\label{sec:VAE}
The variational autoencoder (VAE) \citep{kingma2013auto} belongs to a family of explicit density models known as a directed latent variable model. Whereas a standard autoencoder will learn a compressed representation of the dataset, the VAE extends this by learning parameters of a probability distribution in the latent space from which samples can be drawn. The biggest use for latent variable models are of interesting domain transformations or feature interpolations, sometimes casually referred to as ``morphs''. Various works have demonstrated this effect by blending two distinct timbres or extending the pitch range of a particular instrument \citep{engel2017neural,luo2019learning}, where the associated timbres and pitches were learnt by the model from data in an unsupervised or semi-supervised fashion. For completion, other partially relevant systems that were not tested on raw musical audio but may be of interest include DeepMind's VQ-VAE \citep{van2017neural} that deals mostly with speech and Magenta's MusicVAE \citep{roberts2018hierarchical}) which uses MIDI.   

\begin{figure}[h]
  \centering
     \includegraphics[width=0.74\textwidth]{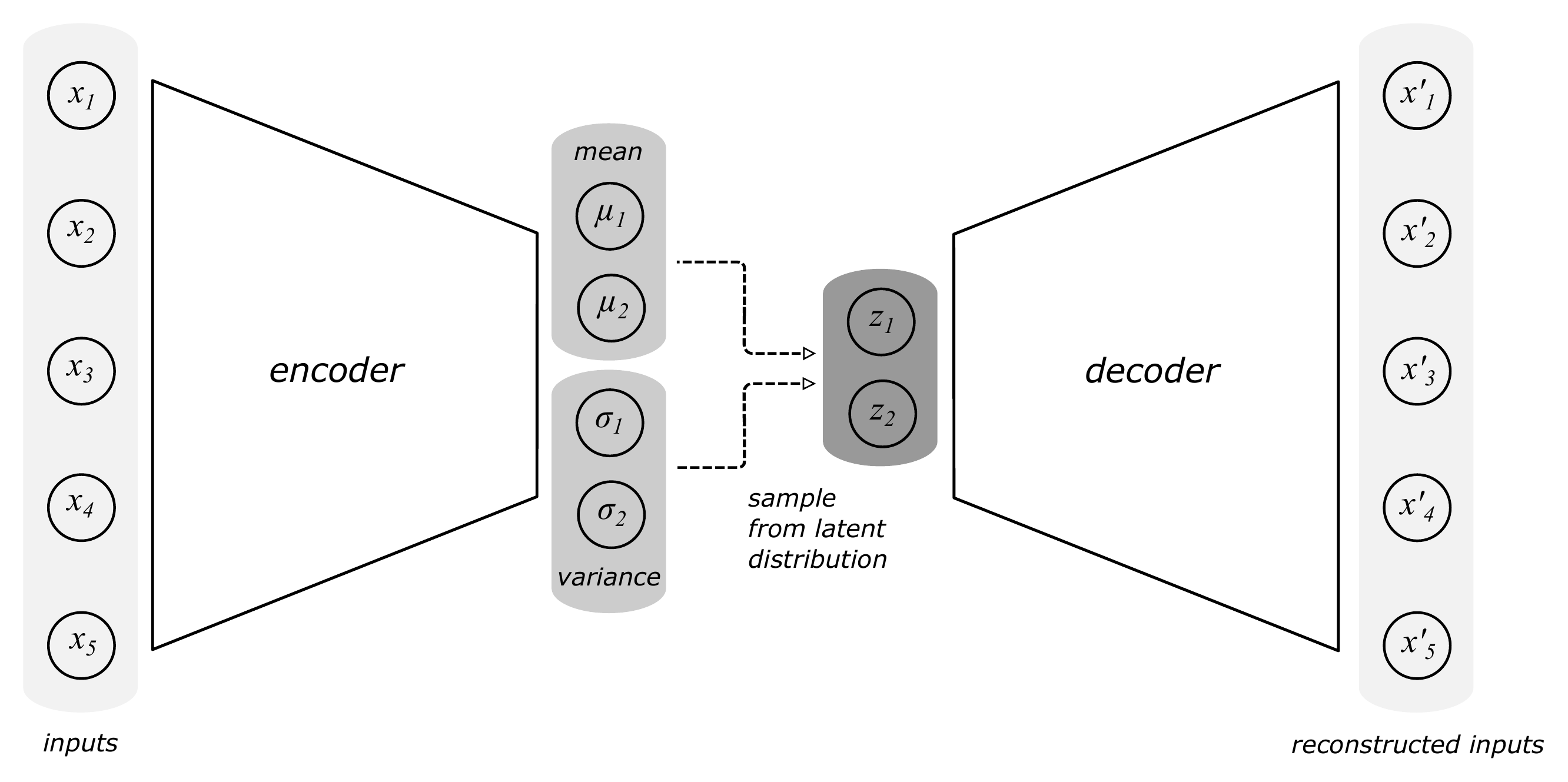}
  \caption{Architecture of a VAE. The encoder processes the input data $X = (x_1,x_2,...)$ and maps them to a Gaussian distributed latent space $q(z|X)$ parametarized by mean $\mu$ and variance $\sigma$. The decoder samples from this latent space using $p(X|z)$ to produce a synthetic output $\hat{X}$. }
  \label{fig:vae}
\end{figure}

The high level architecture of VAEs, shown in Fig. \ref{fig:vae}, is similar to the standard autoencoder -- an encoder network takes inputs and maps them to latent space comprised of latent variables $z$, then a decoder network uses the latent variables to produce an output. The VAE places additional constraints on the form of the latent space and introduces a loss function based on KL divergence for probability distributions jointly trained with the standard autoencoder reconstruction loss.

To motivate the VAE framework, we start with an inference model from Bayes' rule.

$$p(z|X) = \frac{p(X|z)p(z)}{p(X)}$$

Unfortunately, in many cases computing $p(X)$ directly is intractable. To approximate the posterior $p(z|X)$, either a Monte Carlo (sampling) approach would be used, or as in the case of VAE, variational inference. The variational approach circumvents the intractability problem by defining a new distribution $q(z|X)$ and tries to make it as similar to $p(z|X)$ as possible by minimising their KL divergence. By substituting the inference model above into the formula for KL divergence and rearranging, we can derive the objective function that we maximise, known in this context as the variational lower bound $\mathcal{L}$\footnote{For full details on the derivation, a good resource is \citet{doersch2016tutorial}.}.

$$\mathcal{L} = \mathbb{E}_{z \sim q(z|X)}[\log p(X|z)] - \mathcal{D}_{KL}[q(z|X) || p(z)] $$  

The first term on the right-hand side can be interpreted as the reconstruction loss. This naturally lends itself to an autoencoder design, trained to reconstruct its own input. Regularisation is provided by the KL divergence term where the form chosen for $p(z)$ is typically a unit Gaussian distribution. The intuition behind this is to encourage the encoder to distribute all encodings evenly around the center of the latent space instead of clustering them apart. The resultant continuous latent space then allows smooth interpolation between variables during generation by sampling from $p(X|z)$. Musically, a smooth interpolation of the latent space can create smooth audio morphs \citep{Slaney1996} by manipulating these parameters that define the distributions from which different types of audio are drawn during generation.

\subsection{Normalizing Flow Models}
\label{sec:flow}
As we have seen, autoregressive models provide tractable likelihoods but no direct mechanism for learning features. On the other hand, VAEs can learn feature representations but come with intractable marginal likelihoods. Normalizing flows is another family of models that combines the best of both worlds, allowing for both representation learning and tractable likelihood estimation. This means that normalizing flow models have access to latent data features that can be used to condition the generation like the VAE, while possibly learning a more accurate representation of the data distribution. 

Perhaps the most prominent utilisation of normalizing flows from the audio generation literature is Parallel WaveNet \citep{oord2017parallel}. It takes advantage of the parallel generation possible with an Inverse Autoregressive Flow (IAF) \citep{kingma2016improved} that was not previously possible with the purely autoregressive nature of the original. This advancement improved the efficiency over the vanilla WaveNet by a factor of 300, essentially making faster than real-time generation achievable. Aside from Parallel WaveNet, other significant flow models developed specifically for audio generation include ClariNet \citep{ping2018clarinet} and the non-autoregressive WaveGlow \citep{prenger2019waveglow} primarily used for speech.  

The key concept behind normalizing flows is to map simple distributions to more complex distributions using the change of variables technique. We start off with a simple distribution such as a Gaussian for the latent variables $z$ that we aim to transform into a complex distribution to represent the audio output $X$. A single transformation is given by a smooth and invertible function $f$ that can map between $X$ and $z$, such that $X=f(z)$ and $z=f^{-1}(X)$. Since a single transformation may not yield a complex enough distribution, multiple invertible transformations are composed one after another, constructing a ``flow''\footnote{For those interested, in the actual probability space, this transformation has an additional term given by the determinant of a Jacobian matrix derived using change of variables. The final marginal likelihood $p(X)$ is given by $p_{X}(X)=p_{z}(f^{-1}(X)) \bigg| \det \frac{\partial f^{-1}(X)}{\partial X}\bigg|=p_{z}(z) \bigg| \det \frac{\partial f(z)}{\partial z}\bigg|^{-1}$}. Each mapping function in the flow can be parameterised by neural network layers. 

\begin{figure}[h]
  \centering
     \includegraphics[width=0.81\textwidth]{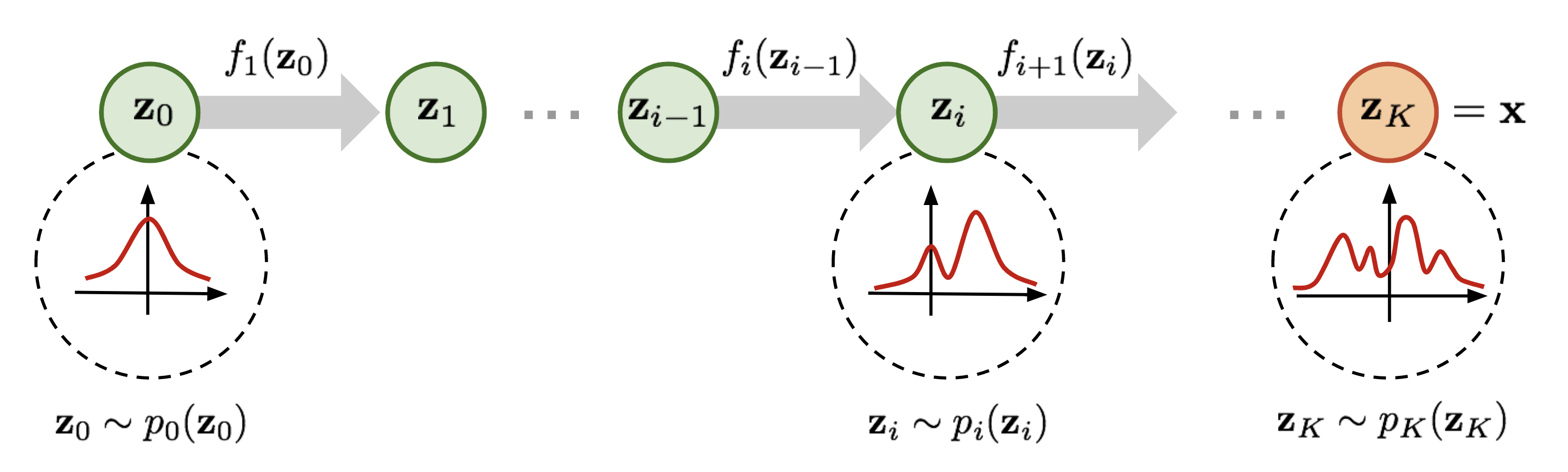}
  \caption{A normalizing flow model transforms a simple distribution shown here as $p_0(z_0)$ to a more complex one $p_k(z_k)$, after $k$ such forward mappings. The resulting output distribution can be taken as the learned distribution of the data. Figure reproduced from \citet{weng2018flow}.}
  \label{fig:flow}
\end{figure}

All the advantages of normalizing flows do come with certain compromises. Unlike the other model families discussed here, normalizing flow models have a much more restrictive set of requirements when it comes to its architecture, namely an invertible mapping function, an easily computable determinant of the Jacobian matrix needed to calculate the likelihood, and an $X$ and $z$ with the same dimensionality.

\subsection{Generative Adversarial Networks}
\label{sec:GAN}
The final family of models to be considered is the generative adversarial network or GAN, a deep learning framework introduced by \citet{goodfellow2014generative}. GANs learn to map random input vectors (typically of much smaller dimension than the data) to data examples in the target domain. They tend to cluster similar output data to neighbourhoods of input values which provides a natural means of navigating among output data with different characteristics. However, in the vanilla GAN, there is no control during training over which regions of input map to which regions of the output space. The user of the model must search and discover the structure after training. 

The development of the GAN architecture was partly motivated to overcome the difficulties inherent in the training of other implicit models\footnote{A primary example is the generative stochastic network \citep{bengio2014deep} that uses Markov chains. Markov chains normally have issues scaling to high dimensional spaces and come with increased computational cost \citep{goodfellow2016nips}.}. Instead it utilises the backpropagation and dropout algorithms that have been so successful with discriminative networks, while introducing a novel training regime based on a \textit{minimax} game.

\begin{figure}[h]
  \centering
     \includegraphics[width=0.62\textwidth]{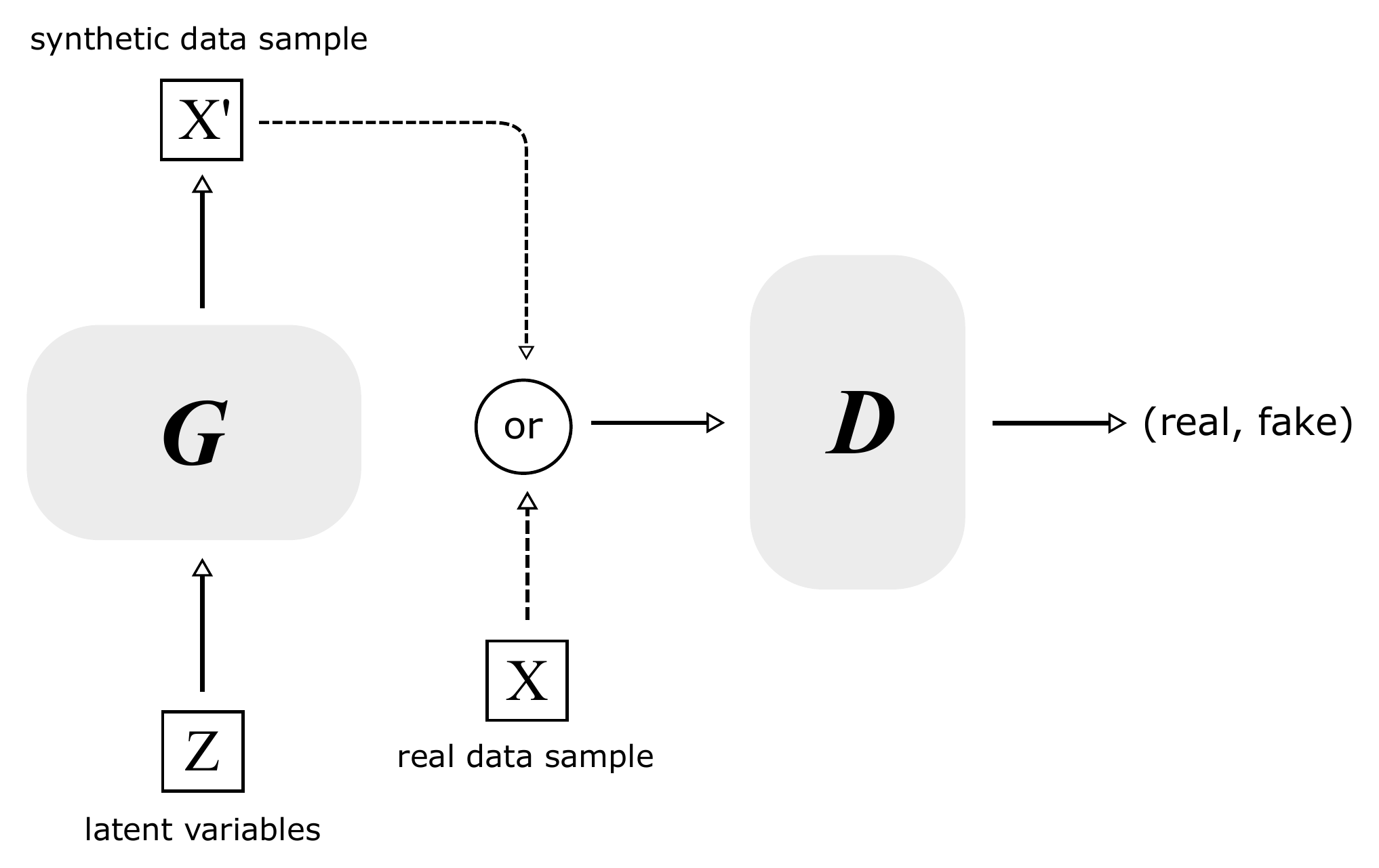}
  \caption{A summary of the adversarial training process of discriminator network $D$ and generator network $G$.}
  \label{fig:gan}
\end{figure}

In a vanilla GAN, the adversarial training pits two separate networks, a discriminator network $D$ and a generator network $G$, against each other (illustrated in Fig. \ref{fig:gan}). We can think of $G$ as a mapping from some input representation space, made up of a prior of latent variables $z$ (traditionally initialised as noise), to some data space $\hat{X}$ (e.g. spectrograms). $D$  maps the input data, be it $X$ or $\hat{X}$, to a categorical label according to whether it thinks the input came from the true data distribution $p(X)$ or the model distribution ${p}(z)$. While $D$ is trained to maximise the probability of distinguishing the real from the synthesized data, $G$ is trained in parallel for the antagonistic objective, that is to fool $D$ by minimising $\log(1-D(G(z)))$. The combined value function $V$ is presented below.

$$G: G(z) \rightarrow \hat{X} $$
$$D: D(X,\hat{X}) \rightarrow (0,1)$$
$$\min_{G} \max_{D} V(G,D) = \mathbb{E}_{X \sim p(X)}[\log D(X)] + \mathbb{E}_{z \sim \hat{p}(z)}[\log(1-D(G(z)))]$$ 

At equilibrium, in an ideal setting (that is difficult to achieve in practice), the generator (re)produces perfectly the true data distribution, which ultimately leads to the discriminator merely randomly guessing the input label, unable to tell the real from the fake. It is important to note that while the discriminator is exposed to both real and modelled data, the generator has no direct access to the training samples, and so has to develop an understanding of the data distribution solely via the error signal provided by the discriminator.



Researchers have pointed out that since the GAN framework does not impose any restriction on the structure of $z$, the generator may use it in a highly entangled way, where individual dimensions do not correspond to semantic features of the data \citep{chen2016infogan}. If various musical instruments are being trained over a range of pitch, for example, it may well be that the input dimensions make it difficult to, say, manipulate timbre holding pitch steady, or to play a scale holding the instrument timbre steady. One way to induce additional structure is to use a conditioning strategy. \citet{mirza2014conditional} present a version that feeds both $G$ and $D$ some extra information, that could be for instance class labels (e.g. pitch and instrument), at input. This setup allows control over the modes of the generated data based on a conditional label. Another extension, InfoGAN \citep{chen2016infogan}, decomposes $z$ into two parts: a noise vector like before, and a latent code that is trained to have high mutual information with the generator distribution. This latent code can be used to discover features in an unsupervised fashion.   

While GANs have seen great success in producing high resolution images, the same level of fidelity has not yet translated to the audio domain, leading to few attempts at using GANs for audio generation. Initial experiments that directly replaced images with spectrograms resulted in low quality samples that can partly be attributed to phase misalignments during the upsampling of the latent variables to achieve the final spectrogram output. GANs do have certain notable advantages over the sample-level generation in autoregressive and some VAE architectures, namely a significant speed-up and a more refined control over global conditioning via its latent variables. Lately there have been more sophisticated efforts for audio such as WaveGAN \citep{donahue2018synthesizing}, VoiceGAN \citep{gao2018voice}, and GANSynth \citep{engel2019gansynth}. The newer works collectively introduce methods that make GAN-based audio generation much more viable than before.

\section{The gift of music: DNN-based synthesizers}
\label{DNN-based audio synthesis systems}

The model families introduced in the preceding section have been used in isolation and sometimes in combination for various audio synthesis tasks. One of the fundamental challenges of generating audio, especially as a raw waveform, is to capture dependencies across long sequences while still maintaining rich timbral information. For perceptually coherent sound this might mean modelling temporal scales across four orders of magnitude, from the sub-milliseconds to tens of seconds. The pioneering models that try to accomplish this objective were autoregressive in nature and led by two competing models, one from Google DeepMind called WaveNet and the other from MILA called SampleRNN. Both focus on modelling at the finest scale possible, that of a single sample, but each use a different network architecture to capture long-term temporal dependencies, namely \textit{dilated convolutions} for WaveNet versus \textit{stacked recurrent units} for SampleRNN. At the time of writing, WaveNet is by far the more popular DNN-based audio synthesis model of the two, and has been incorporated into many other systems as the terminal processing block to map output features to audio. 

\subsection{WaveNet}  

WaveNet's design \citep{van2016wavenet} is rooted in the PixelCNN architecture \citep{van2016conditional} where each step models a joint probability $p(X) = p(x_t|x_1,...,x_{t-1})$ to predict the next audio sample conditioned on a context of $t$ past samples. The architecture consists of a stack of convolutional layers; no pooling layers are present so as to keep the dimensionality between the input and output consistent. 

\begin{figure}[h]
  \centering
     \includegraphics[width=0.77\textwidth]{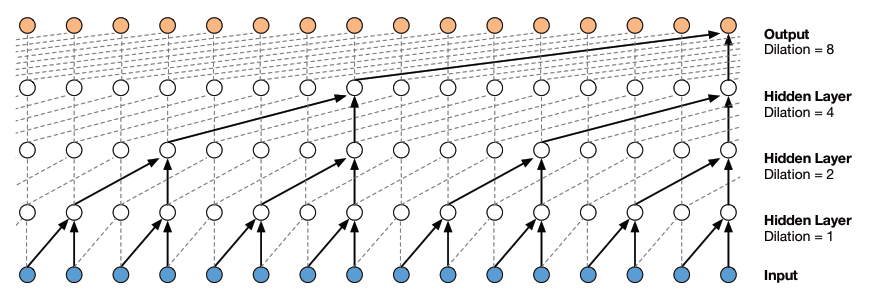}
  \caption{High-level view of the WaveNet architecture showing the dilated causal convolutional layers. Figure reproduced from \citet{van2016wavenet}.}
  \label{fig:wavenet}
\end{figure}

Two main features of WaveNet's convolution operation is that it is \textit{dilated} and \textit{causal}. Dilated convolutional layers expand the extent of the receptive field by skipping time steps, enabling the output to be conditioned on samples that are further in the past than is possible with standard convolutions.  As illustrated in Fig. \ref{fig:wavenet}, each filter takes every $n$th element from the previous layer as input instead of a contiguous section, where $n$ is determined by a hyperparameter called the dilation factor. As the receptive field grows exponentially with depth, fewer layers need to be implemented. In addition, the causal component means that ``future'' samples relative to the current time step do not factor into the next sample prediction.

Alongside the convolutional layers, the same gated activation units as PixelCNN were used as non-linear activation functions, while residual and skip connections \citep{he2016deep} were utilised to improve convergence. van den Oord previously demonstrated that modelling the output as a categorical softmax distribution over quantized possible pixel values worked better than real-valued distributions such as a Gaussian mixture model \citep{oord2016pixel}. For WaveNet, $\mu$--law encoding was first applied to the audio before quantizing to 1 of 256 possible values. For each time step, each of the 256 possible values for the next sample is assigned a probability representing a distribution from which the next sample is drawn. Experiments with WaveNet modelled both music and speech but it is also capable of multi-speaker speech generation when conditioned on speaker identity and text-to-speech when conditioned on text-derived features.

\subsection{SampleRNN} 

\begin{figure}[h]
  \centering
     \includegraphics[width=0.83\textwidth]{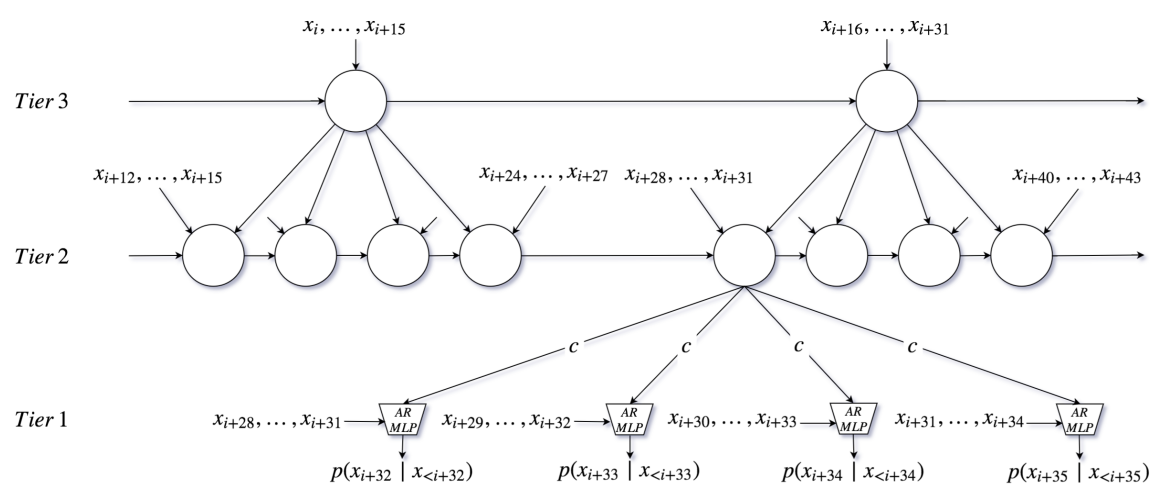}
  \caption{Overview of the SampleRNN architecture. Higher tiers capture increasingly longer time frames which are used to condition the tiers below. Figure reproduced from \citet{mehri2016samplernn}.}
  \label{fig:samplernn}
\end{figure}

Instead of convolutional layers, the SampleRNN architecture \citep{mehri2016samplernn} relies on long short-term memory (LSTM) units/gated recurrent units (GRUs), which are variants of the RNN, operating in a hierarchy of layers over increasingly longer temporal scales. The recurrent modules in each layer summarises the history of its inputs into a conditioning vector that is then passed to the subsequent layer all the way to the lowest level modules made up of multilayer perceptrons, which then combines this information with preceding samples to output a distribution over a sample as seen in Fig. \ref{fig:samplernn}. As with WaveNet, the output is discretized with a softmax.  

Qualitatively, the subjective performances of WaveNet and SampleRNN are very similar. Quantitatively, the SampleRNN authors claim slightly better negative log-likelihood scores for SampleRNN compared to WaveNet on several datasets for speech, non-linguistic human sounds, and music. Nevertheless, the authors concede that their re-implementation of WaveNet may not have been optimal due to a lack of published information regarding hyperparameters.

\section{Only a matter of time: Real-time generation}

WaveNet and SampleRNN significantly improved the quality of audio synthesis over previous parametric and concatenative methods, particularly in terms of the naturalness of its speech synthesis. In the initial experiments, both were trained on not only speech but music databases as well, in particular MagnaTagATune \citep{law2009input} and various solo piano datasets, showing that the models can generalise, to a degree, to different audio domains. These early audio generative models had a huge impact on the deep learning community and moreover, brought attention to the possibilities of audio generation with deep generative networks. With gains in quality, the next step required to actually utilise these models in a production or performance environment was to boost the generative speed. Although the architecture of these autoregressive models is causal, and could thus in theory operate under real-time control, they were dependent on inefficient ancestral sampling resulting in an inherently slow synthesis speed, making the original WaveNet and SampleRNN far from being able to generate audio in real-time. Subsequent years of  development have focused on improving efficiency for speed and for the ability to run with restricted resources on embedded systems or mobile devices while preserving audio quality. 

\subsection{(Faster than) real-time synthesizers}

A direct approach to mitigate the problem was to improve the computational efficiency in these models. Fast WaveNet \citep{paine2016fast} was one of the earlier improvements over the original, which implemented specialised lower-level kernels to store intermediate calculations through caching. This reduced the need for recomputing WaveNet nodes from scratch for each sample generated. Furthermore, the temporal gains acquired by the method scales with the number of layers and could generalise to other autoregressive networks.

Other implementations try to push efficiency further by redesigning the components that significantly contribute to computational time themselves. For WaveRNN \citep{kalchbrenner2018efficient}, this was done by reducing the number of parameters and operations, implementing custom graphics processing unit (GPU) operations, and generating samples in a more parallel manner. The core of WaveRNN is a single-layer RNN followed by two fully-connected layers ending with a softmax output that predicts 16-bit audio samples (in comparison the original WaveNet and many other papers output 8-bit $\mu$-law encoded samples). The state-based operation of an RNN is advantageous in this context as it can perform highly non-linear transformations without requiring the multiple stacked layers present in convolutional architectures like WaveNet, thus reducing the overall number of operations. The bigger contributor to the speed-up nevertheless is the heavy use of low-level code optimisation to overcome memory bandwidth and calculation execution bottlenecks. The resulting WaveRNN implementation can produce 24kHz 16-bit audio at 4$\times$real-time. 

A significantly simpler model in comparison to the original WaveNet is FFTNet \citep{jin2018fftnet} which contains layers that operate like a fast Fourier transform (FFT). Here the input into each layer is split, transformed separately using 1x1 convolutions, then summed. Each step corresponds to a process in the Cooley-Tukey FFT algorithm \citep{cooley1965algorithm}: separating even and odd discrete Fourier transform (DFT) components, calculating the complex exponential part of the DFT, then combining the two DFT halves into the full DFT. It is much faster than the vanilla WaveNet given the simpler architecture, and real-time generation is possible.  

Several studies have sought to reformulate autoregressive models as normalizing flows to skirt around the inherent inefficiency of sample-by-sample autoregressive synthesis networks. These models preserve the general structure of a WaveNet model but redesign the network layers or training procedure to adhere to the restrictions of a flow-based set-up. Compared to the purely autoregressive models, their flow-based counterparts can take better advantage of the massively parallel computation capabilities of a GPU, particularly during the generative phase. In particular, Parallel WaveNet \citep{oord2017parallel} combines two types of flow models, a flow-based WaveNet student model that generates samples and another pre-trained WaveNet teacher model to score the generated samples. Together they are trained by comparing their KL divergence. This teacher-student training scheme is known as \textit{probability density distillation} and has shown to achieve 20$\times$real-time synthesis for 24kHz speech samples albeit at the expense of longer training times. Parallel WaveNet has been successfully deployed in Google's Google Assistant systems.

Notably, Parallel WaveNet had to apply a Monte Carlo method to approximate the KL divergence between the distributions given by the teacher and student models. This may lead to large variances in the gradient calculation during backpropagation due to the sampling procedure which are not ideal for training. ClariNet \citep{ping2018clarinet} introduced a distillation algorithm that instead estimates the KL divergence in closed-form, largely stabilising the training procedure. 

\section{The answer lies within: Interfacing via conditional models}
\label{sec:conditioning}

Notwithstanding their high quality, there are several issues to consider for the DNN-based synthesizers in the preceding section to reliably function as a synthesizer. Despite the special architectures designed to capture long-term dependencies, there is a limit to how much the receptive field of the network can be stretched, and we quickly get into memory bottlenecks with increasing number of layers. Moreover, being autoregressive in nature, each generated sample is only directly conditioned on the immediate previous samples, hence the influence from samples further back in its output history quickly diminishes. Effectively, the synthesized audio is only coherent up to the 10-100ms level. Indeed this is most obvious in unconditional versions of WaveNet that, while getting the characteristic timbre of human speech correct, produces incoherent babble akin to splicing random phonemes. Otherwise when trained on music data made up of multiple instruments, it produces extremely unstructured music that jumps around between different timbres -- hardly a viable model to be used as a synthesizer. 

In the generative task, we aim to learn a function that approximates the true data distribution of the training set via our model, then sample from this approximate distribution to produce new examples. Since the model presumably learns the entire data distribution, sampling from it unbiasedly would consequently yield random pieces of generated data originating from anywhere in the distribution. Instead of generating data from the entire distribution, what we often want in practice is to produce something only from certain subsections of the distribution. One might therefore introduce a sampling algorithm as a means of guiding the output.

Graves in his influential paper on generating sequences with RNNs \citep{graves2013generating}, outlined a few methods to constrain the sampling of the network. One way is to \textit{bias} the model towards churning out more probable elements of the data distribution, which in Graves' example corresponded to more readable handwriting. The sampler is biased at each generative step independently by tuning a parameter called the probability bias $b$ which influences the standard deviation of the probability density of the output. When $b=0$, unbiased sampling is recovered, and as $b \rightarrow \infty$, the variance vanishes so the model outputs an approximation of the mode of the posterior distribution. Similar control parameters that affect the randomness of output samples appear in other works. For Performance RNN \citep{performance-rnn-2017}, a network designed to model polyphonic music with MIDI, a ``temperature'' parameter reduces the randomness of events as it is decreased, making for a more repetitive performance.

\begin{figure}[htb]
  \centering
     \includegraphics[width=0.48\textwidth]{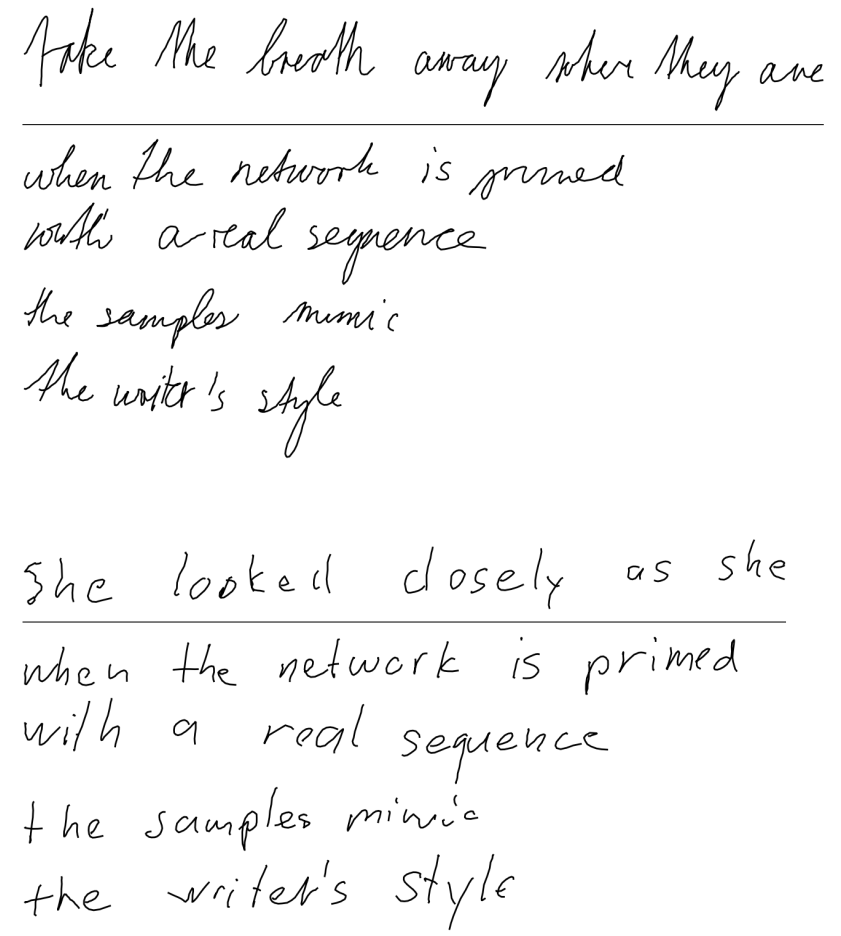}
  \caption{Handwriting examples generated by a LSTM-based network. The top line of each block (underlined) shows the priming sequences drawn from the training set. The following generated text, while never having existed in the training set, retains the handwriting style of the priming sequence. Analogously, generated audio samples would retain the sound characteristics of the priming sequence over a finite number of time steps. Figure reproduced from \citet{graves2013generating}.}
  \label{fig:handwriting}
\end{figure}

While biased sampling introduces one aspect of control, not much can be done by way of choosing the kinds of examples we want to generate. For RNNs in sequence generation, the network can be \textit{primed} by presenting the network with a user-specified sequence that determines the state of the recurrent units before the generation phase begins (when predicted output samples are used as input to predict following samples). Any characteristics embodied by the priming sequence will tend to continue into the generating phase assuming they extended through time in the training data. This is true not just for easily describable characteristics such as pitch, but also for characteristics harder to describe but statistically present (over limited durations) such as ``style'' characteristics. \textit{Priming} allows a higher degree of control over desired aspects of the dataset than bias manipulation. Again referencing Graves' handwriting study, primed sampling allows generation in the style of a particular writer rather than a randomly selected one. As illustrated using handwriting in Fig. \ref{fig:handwriting}, the generated output following the priming sequence can retain the stylistic aspects of the primer.

The priming technique does have several weaknesses. As the RNN's memory capacity has shown to be limited in practice,  for longer sequences, the generated samples tend to ``drift'' away from possessing the characteristics of the priming sequence. \citet{Glover2015} has a sound example of an RNN trained to generate a tone which slowly drifts around the frequency of the priming tone. Also, priming conflates many perceptual dimensions without allowing more fine-tuned control over them independently. In the case of the handwriting example above, the entire style of writing is preserved. However, one cannot individually alter lower-level stylistic aspects like the roundness of letters or the spacing between letters and so forth. This lack of specific control as well as the lack of real-time interaction throughout the generative phase limit the potential use of priming for real-time sound models.


Fortunately, given that a generative model learns the entire joint distribution over a dataset, it is possible not just to alter the sampling procedure but to change the distribution itself by modelling a distribution depending on externally imposed conditions. Mathematically, this means that instead of sampling directly from the model distribution $p_{\theta}(X)$, we sample from a conditional probability $p_{\theta}(X\mid C)$ given the set of conditions $C=\{c1,c2,c3,...\}$. As such, the model output is constrained only to certain subsections of the posterior distribution that are congruent to the conditions imposed. It is often useful to condition on rich side information that correspond to certain high-level features of the dataset. If the model has learned the data well, by specifying certain factors of variation, such as pitch or timbre in the case of music data, one can then contextualise the type of output desired. Overall, the idea of \textit{conditioning} makes for an extremely appealing approach to constructing control parameters required to guide a synthesis model.     

\begin{figure}[h]
  \centering
     \includegraphics[width=0.80\textwidth]{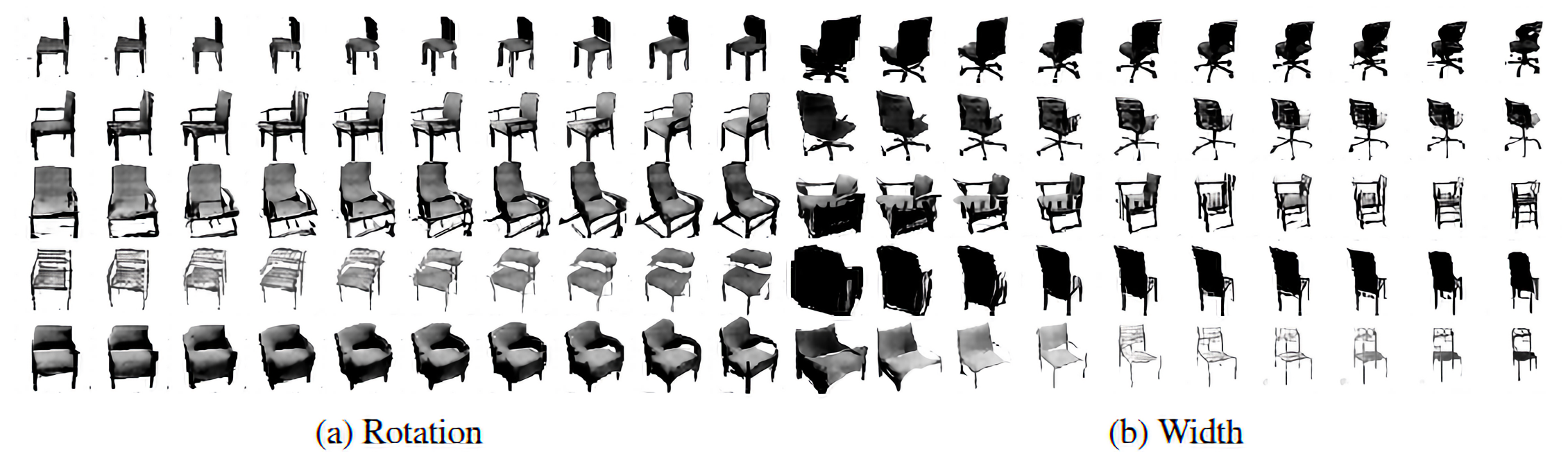}
  \caption{Sampling from the latent space in the InfoGAN model shows how it captures rotation and widths of a chair while preserving its shape. Notably, the learned ``semantic'' concepts are abstracted from the type of chair, meaning it is generalizable across different examples. In the same way, a generative model for audio can potentially learn prosodic or stylistic elements of sounds in an unsupervised fashion which can then be further utilised as conditioning information. For music this could mean having access to latent codes for timbre, pitch, and other useful audio properties, potentially to be used for many creative applications. Figure reproduced from \citet{chen2016infogan}.}
  \label{fig:latent}
\end{figure}

Conditioning can be induced at either a global level (effecting the entire output signal) or a local level (dynamic at every time step) and can come from labels external to the network or from the network's own feature space. Utilising the feature space of the network is especially compelling since on top of reducing the need for laborious labelled data, the network has the ability to discover for itself conditioning parameters during training (see Fig. \ref{fig:latent} and Sec. \ref{sec:latent}). Furthermore, regular or continuous provision of conditioning information presents the network with a strong signal to maintain structure over a longer time span that might otherwise be lost due to limitations on long time-dependencies. \citet{wyse2018real} showed that the pitch drift described earlier in an unconditional RNN is non-existent with a similar network trained conditionally with pitch information. Conditioning frees the network from having to model dependencies across long time scales so that more capacity can be devoted to local structure.     

In essence, conditioning is a way to restrict the probabilistic model only to certain configurations that we care about. In practice, leveraging data to drive both the synthesis and control affordances goes a long way towards addressing some of the issues introduced earlier; conditioning information simultaneously provides an interface for control and restricts the output to the correct set of parameter combinations. From an engineering point of view, transitioning to a conditional model requires two major ingredients: a way to inject supplementary information into the model pipeline during training, followed by a way to expose the same labels again for user control in the generative phase. These developments will be the main focus of the following sections.

\section{Along for the ride: External conditioning} 
\label{sec:external}

A basic music synthesizer interface would typically give the user an option to choose the instrument desired, along with a way to change common musical parameters like pitch and volume. One way to adapt this control structure for a deep generative model is to introduce additional labels which can come in many different forms, including categorical class labels, text, one-hot encoding, and real-valued numbers representing pitch, loudness or duration, depending on the control dimension desired; each corresponding to a musical parameter. The desired combination of labels can be concatenated and fed to the model as a conditioning vector together with the audio samples themselves.

The actual implementation for conditioning may be further broken down into two different methodologies according to whether the labels originate externally with respect to the generative model or discovered internally as part of the unsupervised representation learning process. External labels may be derived directly from raw inputs that are prepared beforehand and annotated by the developer (Sec.\ref{sec:annotated}) or otherwise pre-processed by other models or algorithms to produce the final conditional vector (Sec.\ref{sec:mid}-\ref{sec:long}). Several studies have investigated the viability of these approaches.

\subsection{External conditioning via observed labels} 
\label{sec:annotated}

\begin{figure}[htb]
  \centering
     \includegraphics[width=0.58\textwidth]{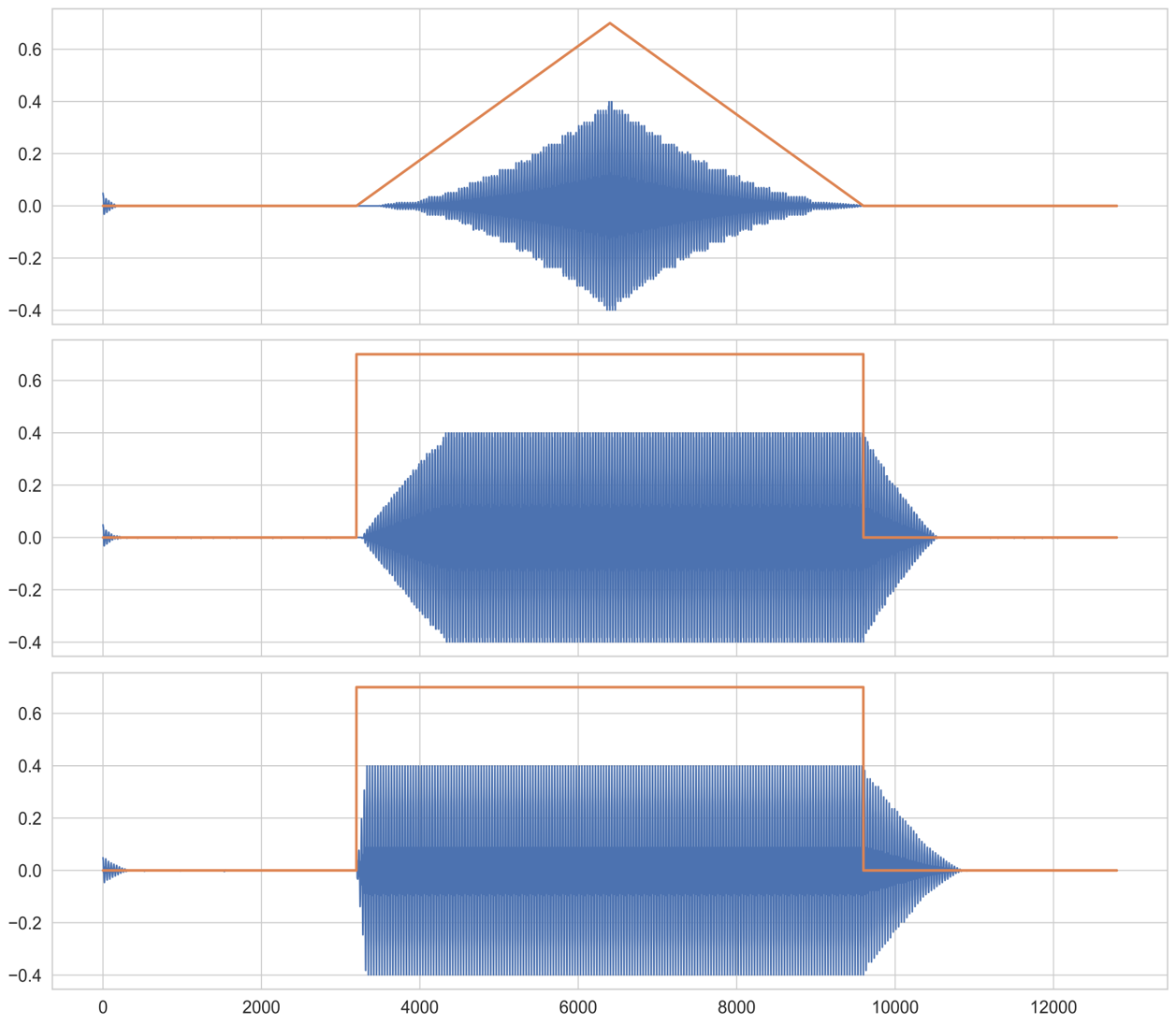}
  \caption{In the top figure, the amplitude of the generated signal tracks a smooth volume change in the conditioning parameter (orange line). However, sudden changes in the value of the volume conditioning parameter trigger transient attacks and decays in the output audio waveform that form over a finite time as they did during the training phase. The middle and bottom figures with their different waveforms and transient characteristics are produced by different settings for the ``instrument ID'' conditioning parameter. Figure reproduced from \citet{wyse2019smc}.}
  \label{fig:transient}
\end{figure}

The extent to which external conditional parameters can influence synthesis was studied by \citet{wyse2018real} via a simple multi-layer RNN model. The network was jointly trained on real-valued parameters indicating pitch, volume, and instrumentation, together with raw audio from the NSynth database \citep{engel2017neural}. During the generative phase, combinations of these parameters were fed as input to the network at every time step, augmenting input from audio predicted from previous steps. The work further highlighted the network's ability to interpolate between conditioning parameter values unseen during training. This was demonstrated through a pitch sweep that interpolated between two trained data points at pitches as much as an octave apart. This work was extended by \citet{wyse2019smc} focusing on the learning and synthesis of acoustic transients (the ``attack'' and ``decay'' segments at the beginning and end of tones). In the experimental setup, transients were modelled as a non-instantaneous response to a change in the volume parameter that resulted in the gradual increase or decrease in the amplitude of the waveform towards the desired loudness level. This demonstrated that conditioning can function to effect the audio generation for a significant period of time following changes in value. This effect is shown in Fig. \ref{fig:transient}.  

Direct conditioning using labels and values based on observed hand-picked characteristics have been shown to allow relatively fine-grained control of the output signal since the conditioning labels are time-aligned and present for every sample of the audio signal. Still, this method requires accurately capturing and aligning labels that may not be readily available or expensive to assemble in terms of time and labour. Labels do not have to be measurable from data but can be perceptually derived as well. However, they must of course be consistent with the data and with each other to be learnable.  

\subsection{External conditioning via intermediate representations} 
\label{sec:mid}

Instead of directly providing conditioning information as raw inputs into the generative model, for certain tasks it may be more effective to further process the information beforehand. Take for example text-to-speech (TTS), a task that can be considered under-constrained given the innumerable ways the same text can be vocalised. On account of text being such a compressed representation compared to audio waveforms, it may be difficult to model audio directly on text. It is usually more prudent to model lower level components such as phonemes, intonation, and duration separately and then provide this information as conditioning parameters for the waveform generation model. In brief, sometimes one or more models separate from the generative network are employed to first map raw inputs to an intermediate representation before feeding this to a generative network. This in effect breaks a complex problem into smaller more manageable tasks, leading to easier convergence during optimisation.

It is not uncommon for these ``pre-processing'' modules to also themselves be deep neural networks. Functionally, these networks play several roles in this context. They convert and integrate various sources of information that may be very different representationally (one-hot encoding, real numbers, text etc.) into a common and probably more unified feature space. They can serve as embeddings which are low-dimensional spaces that high-dimensional information such as audio and text can be mapped onto. Embeddings usually contain meaningful distance relationships between values (i.e. features semantically close together are also close in the embedding space. For example, a human male voice would be closer to a human female voice compared to a bird chirping) and so serve as a more compact and useful representation to train the model on. Finally, they are feature extractors that can tease out any helpful latent variables present in the data to aid in the learning of their task.     

\begin{figure}[htb]
  \centering
     \includegraphics[width=0.76\textwidth]{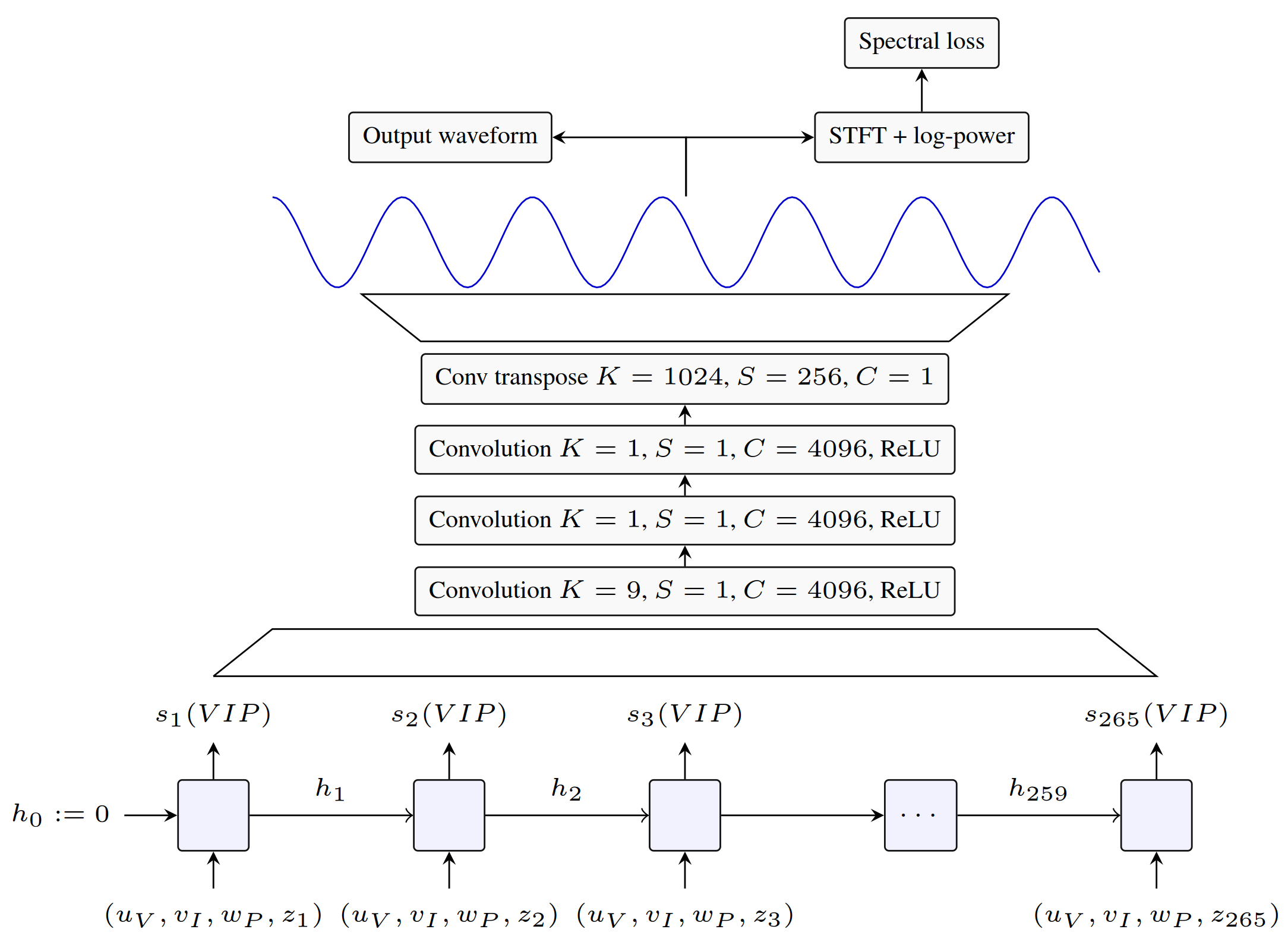}
  \caption{Overview of the SING architecture. The LSTM generator at the bottom produces an intermediate sequential representation from velocity, instrument, pitch, and time embeddings ($u_V$, $v_I$, $w_P$ and $z_{1-256}$ respectively) The sequence is then fed through a convolutional decoder to synthesize a waveform. Figure reproduced from \citet{defossez2018sing}.}
  \label{fig:sing}
\end{figure}

A music model that does this multi-step feature extraction to obtain conditioning information for a generative model is SING (Symbol-to-Instrument Neural Generator) \citep{defossez2018sing}. SING aims to synthesize audio a frame at a time (consisting of 1024 samples), making it arguably more efficient than the sample-level autoregressive models. The main architecture consists of two parts: an LSTM-based sequence generator and a convolutional decoder (see Fig. \ref{fig:sing}). Unlike the autoregressive networks that require prior audio samples to generate what follows, SING's sequence generator takes in embedded values corresponding to audio parameters of velocity, pitch, and instrument plus a component to quantify time dependency, without actual audio samples as input. The generated sequence, which can be regarded as an intermediate feature representation, is then processed and upsampled through several convolutional layers to obtain the final real-valued waveform. 

Training such a multi-component network is usually not a trivial undertaking. Each module in SING must be separately trained before fine-tuning the entire model. For instance, the convolutional decoder is initially pre-trained together with an encoder as an autoencoder. The encoder was subsequently dropped since it was no longer required for the generative task. 

A compelling contribution of the paper is the adoption of spectral loss, which was shown to be superior to mean square error (MSE) commonly used when comparing real-valued objects. The authors surmised that this could be due to errors in phase reconstructions when training with the latter. During generation a model chooses implicitly a phase value that could take any number between [0, 2$\pi$], altering the output in a somewhat random fashion unless it has learnt all possible phase and frequency combinations. This unpredictability makes a MSE comparison with the ground truth in the time domain uninformative and unlikely to train the model well. In comparison, the spectral loss is calculated using the magnitude spectra of the signal, discarding phase information for both the generated waveform and the target waveform. Hence, the model is free to choose a canonical phase by which to base the generation on instead of devoting capacity to learn all possible phase combinations.    

Various extensions to WaveNet and SampleRNN also employ one or more external models to convert raw inputs to an intermediate representation \citep{sotelo2017char2wav,arik2017deep}. Many of these were purposed for TTS tasks and were pre-trained on existing TTS systems to extract vocoder features. In a similar spirit to SING, Tacotron \citep{wang2017tacotron} also implements RNN-based networks to sequentially predict mel spectrograms for speech, contingent upon embeddings for speaker, text, and reference spectrograms. The original version of Tacotron uses the Griffin-Lim algorithm \citep{GriffinLim84} to invert the output spectrograms but this was later replaced with a WaveNet in follow up work \citep{shen2017natural}. The Tacotron team in a series of papers \citep{wang2017uncovering,wang2018style,skerry2018towards} displayed some ways for conditioning to be utilised more creatively that may be more in line with how music synthesis systems would later be designed. ``Style tokens'' were introduced to extract independent prosodic styles from training data within a Tacotron model. Style tokens comprise of a bank of embeddings shared across training samples. The embeddings were trained without any explicit labels (i.e. unsupervised) but are able to generate interpretable parameters that can control synthesis in interesting ways. 10 style tokens where used for experiments, enough to adequately present a small but rich variety of prosodies. They were found to be able to generalise beyond learned inputs, generating audio in the corresponding embedded style even on unseen text. The group's research in style tokens culminates in an enhanced Tacotron with explicit prosody controls allowing changing prosodic style while preserving speaker identity.

\subsection{Capturing long-term structure in music}
\label{sec:long}
As highlighted by \citet{dieleman2018challenge}, autoregressive models excel at generating raw waveforms of speech, whereas when applied to music, they tend to be biased towards capturing local structure at the expense of modelling long-range correlations. One way to alleviate this problem is to divest the modelling of long-term correlations to a conditioning signal. This relieves the network from having to model signal structure beyond a few hundred milliseconds and dedicate more modelling capacity to localised patterns. Analogous to how text is used to condition a neural synthesizer for TTS, several papers have sought the use of MIDI notes or piano rolls as a way of directing a musical signal. This takes advantage of the fact that symbolic representations are easier to model over the long term, while still preserving the expressivity and richness of raw audio models. 

To that end, \citet{manzelli2018conditioning} provided MIDI pitch and timing as conditioning for a WaveNet model. Rather than feeding these values by hand, a second generative network was used to compose MIDI sequences. This second network essentially takes over the modelling of long-range correlations, with the final synthesized audio output shown to follow the timing and pitch of the MIDI notes fairly closely. This idea was further expanded by \citet{hawthorne2018enabling} in their Wave2Midi2Wave system that adds an encoder network to transcribe raw audio to MIDI, plus a state-of-the-art transformer network \citep{huang2018improved} to generate MIDI sequences instead of a LSTM-based one used by Manzelli et al. The paper reported that there was not a statistically significant difference in ratings between real recordings and those generated by the model conditioned on a MIDI test set. Ratings were given by human assessors in a pair-wise blind test according to which clip they thought sounded more like it came from a person playing a real piano\footnote{The system was trained on the MAESTRO dataset that contains over a week of paired audio and MIDI recordings from nine years of International Piano-e-Competition events.}. 

Conditioning can be used for many other applications in music besides modelling long-term structure. In the original WaveNet paper, the authors mooted conditioning the model with a set of tags specifying musical characteristics like genre and instrument identity essentially as a way to alter timbre but did not go in to detail about these findings. This line of research has since been taken up by others primarily through a latent variable model approach where the latents are jointly learned as part of training the generative model instead of depending on external models or direct observable labels. Open AI's Jukebox \citep{Dhariwal2020} uses several VQ-VAEs \citep{van2017neural} in parallel to produce quantized embeddings at different temporal resolutions to help maintain coherence over long time spans. Training is tuned such that each stage encodes information at different levels of abstraction, with the topmost level capturing high-level semantics like melody, while the middle level captures more local features like timbre. Decoding is likewise carried out progressively over multi-timescale stages and uses a combination of autoregressive transformers and WaveNet-style dilated convolutions to model embeddings over time and upsample them to retrieve audio samples. The system was trained on 1.2 million songs, and is able to generate musically coherent samples in a wide variety of styles at timescales of minutes at a CD-quality sample rate of 44.1kHz. Conditioning information on artists and genre allow for high-level control over musical style. Further, conditioning on text allows coherent lyrics to be produced accompanied by the music.

\section{Beneath the surface: Latent variable models of music}
\label{sec:latent}


There may be times when the audio training dataset is too cumbersome for hand-labelling; meta-data may be incomplete, or it may be difficult to process features equitably across the entire dataset. In cases such as these, we can instead pass the burden of extracting useful features to the model itself. With a latent variable model such as the VAE, feature extraction can be carried out internally in an unsupervised or semi-supervised manner. While many other deep learning model variants also learn features to aid its training objective, a latent variable model goes a step further to impose structure on the feature space and adds a generative model that samples from this space. This process entails the discovery of underlying latent factors of variation in the dataset as part of the training procedure (and often encouraged by the objective function itself), in comparison to a more direct mapping from inputs to a feature space akin to a pre-processing step for many of the conditional models discussed in the prior section. On the other hand, compared to the previous models discussed, particularly the autoregressive models, latent variable models are often harder to train.

Having these useful properties, latent variable models are often utilised for two functions that are musically relevant. Firstly, features can be ``disentangled'' from the source material. Take for example a dataset made up of audio clips of several instruments each playing a range of notes. Each piece of data can be said to have an intrinsic timbre and pitch. During training, the model attempts to decouple these two features to conceptualise a characteristic timbre for each intrument as well as separate the different pitch levels in a way that is abstracted from any particular audio sample. If successfully modelled, one can then treat timbre and pitch as independent features to possibly be used as conditioning for the synthesis. After disentanglement, the assorted features can be mixed in creative ways. \textit{Timbre transfer} is the process of selecting a particular timbre and composing it with a pitch not found for that instrument in the dataset or even outside of the instrument's physical range in real life, creating a novel sound. The concept has been demonstrated in principle by \citet{hung2018learning} with separate VAE variants, one utilising distinct encoders for pitch and timbre and the other employing skip connections to process pitch on a different pathway than that of timbre. Both were trained on constant-Q transformed (CQT) audio to output a piano-roll-like symbolic representation. This work was later extended by \citet{luo2019learning} using a more general mel-spectrogram representation. 

Being able disentangle features and construct latent spaces have allowed researches to extend seminal work on timbre spaces reliant on perceptual ``dissimilarity'' scores from human subjects then analysed by Multi-Dimensional Scaling (MDS) \citep{grey1977multidimensional,wessel1979timbre}. Latent variable models admit organising a timbre space in a more objective manner with meaningful distance relationships and so can be treated as an embedding. The unsupervised organisation of the timbre space was studied by \citet{kim2019neural} in relation to their Mel2Mel architecture which predicts mel spectrograms for conditioning of a WaveNet in a similar way to Tacotron. This work in particular used FiLM (feature wise linear modulation) layers \citep{perez2018film} to learn temporal and spectral envelope features from the instrument embeddings that drives the timbre during synthesis. Visualisations of the learned instrument embedding space in Mel2Mel indicated separation along the lines of spectral centroid and mean energy. Although this model seemed to have disentangled timbre in a perceptually meaningful fashion, being unsupervised, there is no guarantee that this will always be the case. In many cases individual latent dimensions may not actually embody or correspond to any clear-cut audio characteristic. This is unlike the external conditioning shown in Sec. \ref{sec:annotated} that, by virtue of being hand-prepared, ensures it can be used as intended during generation. To overcome this, \citet{esling2018generative} incorporated the timbre space from the traditional MDS studies by using it as the prior distribution to train a VAE model, thus guaranteeing some perceptual correspondence. The learned space is topologically similar to the MDS space but with added advantages by virtue of being a VAE, including being continuous and generalizable.     

The benefit of a structured feature space is having a principled way of navigating its topology. This is especially significant for a non-ordinal parameters like timbre which, by itself, has no natural arrangement. Since we now have a continuous embedding with meaningful distances, interpolation, the second major use latent variable models, becomes possible. Rather than sampling from the centre of mass of known features in the feature space as is done when selecting a particular timbre and a particular pitch both present in the training dataset, it is possible to instead choose a point somewhere in between two distinct features. Smooth interpolation allows the model to blend the characteristics of surrounding features when sampling from an interpolated value. In this way, sounds can be synthesized that generalises beyond the training dataset. For instance, a point in the middle of two known pitches in the feature space can be selected which would roughly correspond to a pitch halfway between two pitches. Perhaps more interesting is the possibility of blending timbres in a perceptually integrated way, quite distinct from the superficial ``mixing'' of audio signals from distinct sources. 

One of the first music synthesis systems to display such use was set out in \citet{engel2017neural}. Here the encoder side of the VAE learns a mapping from raw instrument tones to an embedding layer that is connected to a WaveNet decoder. Their encoder is built as a 30-layer nonlinear residual network of dilated convolutions follow by 1x1 convolutions and average pooling. The embedding result is sequential, with separate temporal and channel dimensions whose resolution is dependent on the pooling stride. Whilst keeping the size of the embedding constant, the stride is tuned for a trade-off between temporal resolution and embedding expressivity. Experiments showed that the embedding space spans the range of timbre and dynamics present in the dataset. For generation, the latent vectors are upsampled in time and provided to the autoregressive model as conditioning parameters. Interpolation between two timbers is thus between two targets that are each changing in time. Interpolation between these factors does result in a perceptual melding of different instruments creating novel sounds. On the other hand, trying to generate multiple pitches from a single embedding preserving timbre and dynamics by conditioning on pitch during training was less successful. The authors postulated that this may have been due to unsatisfactory disentanglement between pitch and timbre.



\section{Build me up, break me down: Audio synthesis with GANs} 

Up to now the generative approaches discussed have been based around likelihood models including autoregressive models, VAEs, normalizing flow models, or a combination of them. Unlike these other models families, GANs are not trained using maximum likelihood. A GAN training objective is instead set up to determine whether or not a set of samples from two distributions are in fact from the same distribution -- a procedure known as the \textit{two-sample test}. Despite the difficulty in training GANs\footnote{Early GAN models proved to be unstable to train, and a vast amount of GAN literature has been dedicated towards improving its training. Some commonly identified problems include a failure to converge \citep{salimans2016improved,mescheder2018training}, vanishing gradient, that is the discriminator reaching its optimum too quickly leading to a stagnation of generator updates \citep{arjovsky2017towards}, mode collapse, where the generator consistently produces similar output regardless of input, leading to problems with diversity \citep{salimans2016improved,metz2016unrolled}, and an unclear stopping criteria, with the generator and discriminator losses often oscillating. Although these key issues have largely been ironed out, active research is still being carried out to improve the overall efficacy of GANs, including reducing the trade-off between quality and diversity \citep{bang2018improved}, and much hyperparameter tuning is still required.} they have shown outstanding success in the vision domain and remains the dominant model family for image generation.   

Several research groups have attempted to apply the GAN framework to audio. The historical development of GAN models for audio was as much a search for representations that work well with GANs as they are of advancements in the network architecture. The naive approach of replacing images with an image-like representation of audio i.e. the spectrogram while retaining most of the vision informed architecture resulted in extremely noisy generated samples. GAN-based audio models have come a long way since then. They operate at a much faster rate than the more prevalent autoregressive models, requiring a single forward pass for generation, and so fall well within the real-time requirement of a music synthesizer.

One of the early relatively successful audio GAN models was WaveGAN \citep{donahue2018synthesizing}. It works in the time-domain and utilises 1-dimensional convolutional filters, with an upsampling strategy based on DCGAN \citep{radford2015unsupervised} but with new procedures such as \textit{phase shuffling} to take care of aliasing artifacts created by convolutional fiters and strides. WaveGAN was the better of two similar GAN architectures presented in the paper, the other of which operated on spectrograms. Evaluation was measured through the \textit{inception score} \citep{salimans2016improved} and human judges. The authors report that WaveGAN has not reached the level of fidelity of the best autoregressive models, but is orders of magnitude faster as it can take advantage of the inherent parallelism of GANs to generate hundreds of times faster than real-time. Of course, despite its computational speed, it is still not real-time interactive because the architecture is structured to generate large chunks of audio at a time -- one second in this case. Furthermore, the parametric input to a WaveGAN (like DCGAN) is a 100-dimensional ``latent vector'' which is trained by mapping noise to sound. The input parameter space of the network must thus be searched after training in order to find desired output sounds. Since similar sounds tend to cluster in the parameter space, a typical search strategy is to search coarsely for the desired ``class'' (e.g. drum type) and then in a neighborhood to fine-tune the audio output. A separate strategy would be necessary to map a desired parametric control space consisting of say, instrument ID and some timbral characteristics to the control parameters of the GAN. This is similar to strategies for mapping gesture to parameters for fixed synthesizer architectures discussed above. 

GANSynth \citep{engel2019gansynth} is, at the time of this writing, the state-of-the-art in several aspects of sound modelling. Like WaveGAN, it is much faster than the autoregressive models, although it suffers the same limitations concerning real-time interaction. Several innovations were introduced that have resulted in very high quality audio as well as the ability to intuitively manipulate audio the parameter space for specific audio characteristics. The model is trained on the NSynth dataset of 4-second musical instrument notes. Thus like WaveGAN, global structure is encoded in the latent parameter space because each input generates long audio sequences (4 seconds for GANSynth). 

GANSynth generates a spectral representation of audio, but one that encodes both the magnitude and the phase of the signal. Furthermore, it is actually the derivative of phase which is coded, referred to as ``instantaneous frequency''. This turns out to make a significant difference in audio quality. A second innovation is the augmentation of the latent vector input with a one-hot vector representation of the musical pitch of the data. After training, latent vectors encode timbre which to a very rough approximation are consistent across pitch for a given instrument, while the pitch component can be used for playing note sequences on a fixed instrument. Interpolation between points in the latent vector space also works effectively.

\section{A change of seasons: Music translation}
\label{style}

There is yet another way to use generative models that opens up many possibilities musically. What we have discussed so far entails sampling from the modelled distribution to synthesize novel audio sequences. During this generative phase, the model maps conditional parameters to output audio which is consistent with these parameters. Instead of conditioning on relatively low-level parameters like pitch or volume, an alternative is to provide actual audio as conditioning. During the training phase, if the input audio and the target audio is different, what the generative model does in effect is to find some transformation function between the two.    

\begin{figure}[htb]
  \centering
     \includegraphics[width=0.60\textwidth]{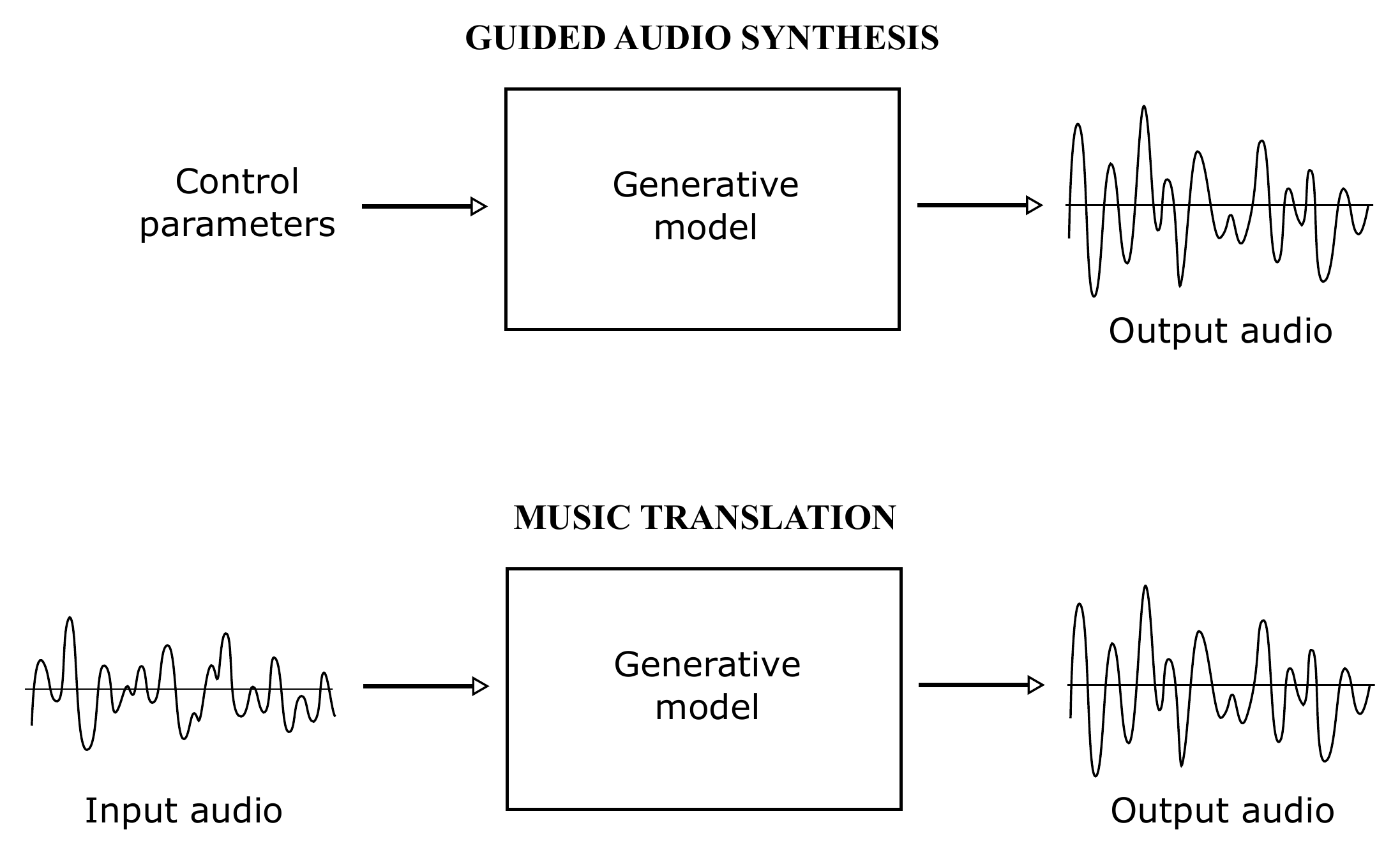}
  \caption{While in the standard guided synthesis task we are interested in generating sound from a set of parameters, for music translation we want to map the sound of one musical domain to another, retaining some audio characteristics of the input domain (e.g. pitch, duration, rhythm) while taking on some other characteristics of the target domain (e.g. timbre).}
  \label{fig:translate}
\end{figure} 

The goal of \textit{music translation} is a domain transfer, where the hope is to imbue the input audio with some characteristics of the target audio while retaining its general structure. During synthesis, this translation should be able to generalise to new waveforms belonging to the original domain. For images, there is a vast amount of recent literature on ``style transfer'' \citep{gatys2015neural} where the desired property to transfer is broadly the artistic style of the target, but preserving the ``content'' or global structure and arrangement of objects in the original image. A more general formulation of this objective can be seen in applications to CycleGAN \citep{zhu2017unpaired} where the model maps sets of images from one domain to another (e.g. photographs to paintings, horses to zebras etc.). We might imagine doing something similar with music and indeed several systems have been developed for such use cases. 

Timbre is one obvious candidate for an audio property to transfer. Timbretron \citep{huang2018timbretron} seeks to do this via a model similar to CycleGAN, working with a CQT representation within the GAN architecture, then retrieving audio through a WaveNet. The training dataset consisted of real recordings of piano, flute, violin, and harpsichord sounds, and the task was to translate one timbre to another. Moreover, to be effective, the system must transform a given audio input so that the output is still perceptible as the same basic musical piece, but recognisably played with the target instrument. The response was broadly positive in those aspects with human evaluators of the system, therefore showing definitive proof-of-concept for music translation. 

The domains of interest can also be more loosely defined. \citet{mor2018universal} presented a ``universal music translation network'' that is able to translate music across instruments, and to a degree, genres and styles. Their novel architecture consists of a single, universal encoder shared across all inputs connected to multiple WaveNets that each decode for a separate musical domain. In addition, a domain confusion network similar to that put forward by \citet{ganin2016domain} was employed during training to ensure domain-specific information was not encoded. Their method displayed an impressive ability to execute high level transformations such as converting a Bach cantata sung by an opera to a solo piano in the style of Beethoven. The domain mapping however has to be decided beforehand as only the desired output decoder (out of the multiple trained WaveNets) is to be exposed to the input during generation. Hence, the user cannot change transformations in real-time. While switching domain transformations on the fly remains an issue, the generative speed itself have been boosted tremendously in later work that replaces the WaveNet decoder with a GAN model \citep{kumar2019melgan}.

Rather than translating across whole instruments or styles which may prove to be challenging and not completely convincing by current methods, one could possibly design for smaller transformation tasks and view the generative model as a complex non-linear filter. \citet{damskagg2019real} modelled several famous distortion guitar pedals like the Boss DS-1 (distortion), Ibanez Tube Screamer (overdrive), and Electro-Harmonix Big Muff Pi (fuzz) with a modified WaveNet architecture. The final deployed model was able to process a clean guitar input signal in real-time on a standard desktop computer, effectively acting as an actual distortion circuit and sounding very close to the actual pedals. Although the mapping from clean sound to distortion is now fixed without the ability to tune further, future work might add the parameters of the original guitar pedals like tone control or gain as conditioning variables to fully emulate their capabilities.

\section{Discussion and conclusion}
\label{sec:discussion}

Machine learning approaches to audio processing in general and sound modelling in particular are quickly gaining prominence and replacing previous techniques based on manual feature selection and coding. Deep learning architectures offer new approaches and new ways to think about perennial sound modelling issues such as designing affordances for expressive manipulation during real-time performance. Hybrid architectures that draw on signal analysis as part of an overal sound modelling strategy are becoming more common. Variational autoencoders, for example, are exploited to discover low-dimensional representations of the data space that can then be used for external control for generation. 

Great strides have been made in capturing complexity and naturalness that is hard to code for manually. Automatic discovery of structure in data combined with specific techniques for regularizing representations have led to intuitive ways of navigating through a space of sounds even across regions where there is no training data. Using strategies for training generative models with conditional distributions is a very musically intuitive way to design interfaces. This methodology shifts the scope of the control space primarily to the training data, or specifically factors of variation in the data (latent or otherwise), from what used to be defined as a by-product of the synthesis algorithm or a design decision for an interface. It is a new way of thinking about how to construct such control affordances in comparison to traditional synthesis techniques. 

Perhaps flexibility is the greatest strength of the deep learning paradigm; models can be reused or redesigned for a wide variety of sound classes. Different model families can be combined to extend their capabilities such as using a VAE to discover latent features that can be passed to an autoregressive generator, or using a normalizing flow model to speed up a standard autoregressive network. This opens up the possibility of attempting more complicated tasks that were not previously possible such as expressive TTS for speech, or real-time timbre transfer for music.

Outstanding issues remain. The best-sounding systems are still very computationally expensive. This makes them challenging to port to ubiquitous low-cost devices. Training typically depends on many hours of data. Data must capture not only sound, but control factors that can be difficult to obtain such as air pressure inside a mouth for a wind instrument model. The amount of data necessary for creating good models can take days to train even on the best GPU platforms which is both time-consuming and expensive. Learning and synthesizing more complex sounds including multi-track, multi-instrument audio is still largely beyond current models. Different types of models have different advantages (VAEs for structuring data spaces, GANs for global parameterization, autoregressive models for the causality necessary for real-time synthesis etc.) but we have yet to see a deep learning system make its way in to general usage for musical audio synthesis the way they have in, for example, the domain of speech synthesis. With the rate that these technologies are being developed, it seems certain to happen in the near future, perhaps even by the time you are reading this manuscript.

\section*{Acknowledgements}

This research was supported by a Singapore MOE Tier 2 grant, "Learning Generative Recurrent Neural Networks,” and by an NVIDIA Corporation Academic Programs GPU grant.

\clearpage
\bibliography{references}  

\end{document}